\documentclass[english]{article}
\usepackage[utf8]{inputenc}
\usepackage[T1]{fontenc}
\usepackage{amsmath}
\usepackage{graphicx}
\usepackage{fancyhdr}
\usepackage{float}
\usepackage{subcaption}
\usepackage{adjustbox}
\usepackage{hyperref}
\newcommand{\scidatalogo}{\includegraphics[height=36pt]{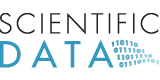}}
\newcommand{\overleaflogo}{\includegraphics[height=36pt]{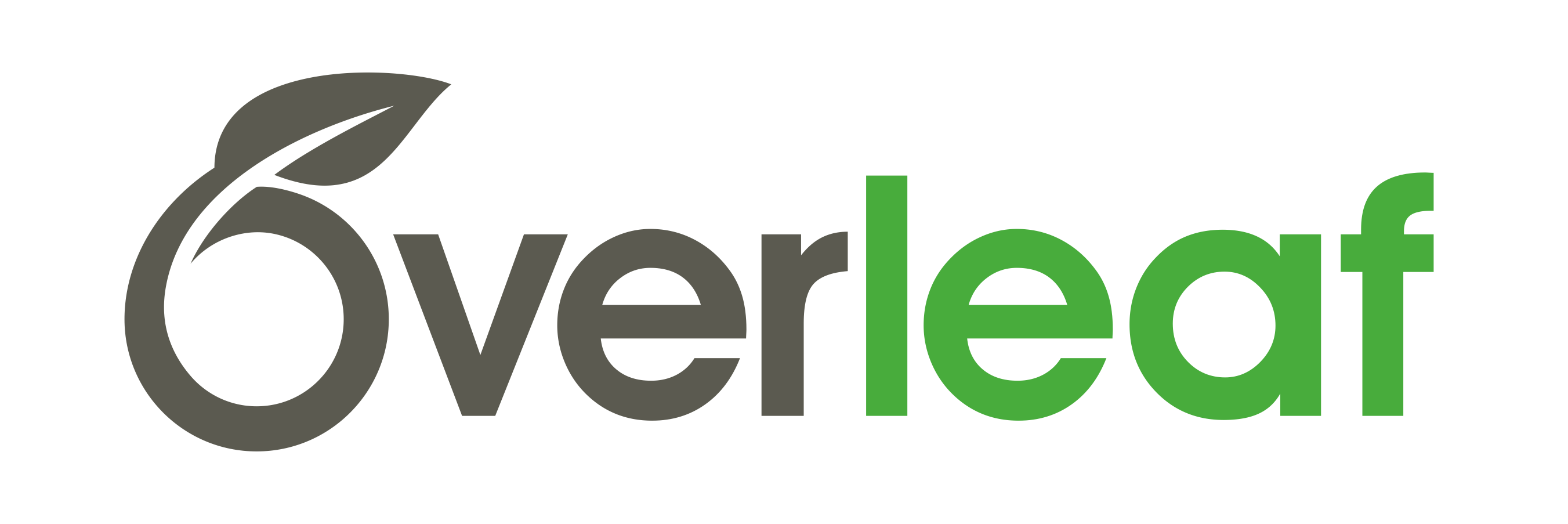}}
\usepackage[dvipsnames]{xcolor}
\usepackage{soul}

\begin{document}

\title{Home-to-school pedestrian mobility GPS data from a citizen science experiment in the Barcelona area}

\author{Ferran Larroya\textsuperscript{1,2}, Ofelia D\'iaz\textsuperscript{3}, Oleguer Sagarra\textsuperscript{3}, Pol Colomer Sim\'on\textsuperscript{3}, \\
Salva Ferr\'e\textsuperscript{4}, Esteban Moro \textsuperscript{5,6}, Josep Perell\'o\textsuperscript{1,2,{*}}}

\maketitle
\thispagestyle{fancy}

1. OpenSystems, Departament de F\'isica de la Mat\`eria Condensada, Universitat de Barcelona, Martí i Franquès, 1, 08028 Barcelona, Catalonia, Spain 

2. Universitat de Barcelona Institute of Complex Systems, Catalonia, Spain

3. Dribia Data Research, Llacuna, 162, 08018 Barcelona, Catalonia, Spain

4. Eduscopi, Esgl\`esia, 69, 08901 L'Hospitalet de Llobregat, Catalonia, Spain

5. MIT Connection Science, Massachusetts Institute of Technology, 02139 Cambridge, MA, USA. 

6. Department of Mathematics and GISC, Universidad Carlos III de Madrid, Madrid, Legan\'es, Spain.

{*}corresponding author(s):
Josep Perell\'o (josep.perello@ub.edu)
\begin{abstract}
The analysis of pedestrian GPS datasets is fundamental to further advance on the study and the design of walkable cities. The highest resolution GPS data can characterize micro-mobility patterns and pedestrians' micro-motives in relation to a small-scale urban context. Purposed-based recurrent mobility data inside people's neighbourhoods is an important source in these sorts of studies. However, micro-mobility around people's homes is generally unavailable, and if data exists, it is generally not shareable often due to privacy issues. Citizen science and its public involvement practices in scientific research are valid options to circumvent these challenges and provide meaningful datasets for walkable cities. The study presents GPS records from single-day home-to-school pedestrian mobility of 10 schools in the Barcelona Metropolitan area (Spain). The research provides pedestrian mobility from an age-homogeneous group of people. The study shares processed records with specific filtering, cleaning, and interpolation procedures that can facilitate and accelerate data usage. Citizen science practices during the whole research process are reported to offer a complete perspective of the data collected.
\end{abstract}

\section*{Background \& Summary}

Intertwined with sustainability, social justice, and climate change, mobility has moved to the middle of policy-makers' agendas in the last years \cite{weforum}. In this context, motorized traffic is considered to be one of the main problems our cities face today as it leads to a wide range of social and socio-environmental consequences. Cities are thus taking several actions to favour pedestrians' environments and promote sustainable mobility and liveability \cite{Eggiman2022}. These efforts also motivate a better understanding of urban mobility broadly and more specifically pedestrians' mobility because issues of walkability and pedestrian advocacy are at the heart of this theme \cite{Rhoads2021}. Much research has been done to understand multi-modal transportation in our cities, especially in the last ten years, due to the availability of digital traces \cite{timegeo}. However, although digital traces work well to investigate large distance movements or origin-destination flows (typical public transportation or vehicles) \cite{cardata}, they are less precise in investigating the local micro-behaviours around people's homes or in pedestrian areas \cite{Hunter2021,Fan2023,Yang2023}. There is an increasing need to get more precise data to better learn about the micro-motives of what makes an area more suitable for walking, how different spaces are used to walk, or what activities are more impacted by walking accessibility. 

In this sense, participatory research initiatives like citizen science \cite{Vohland2021} in which people share their own data, and their pedestrian behaviour is an unmatched source of understanding of human mobility patterns \cite{kapenekakis}. Citizen science broadly refers to the active engagement of the general public in scientific research tasks \cite{Vohland2021}. Among many other aspects related to responsible research practices or scientific literacy, citizen science can offer a powerful way to collect data that otherwise can be difficult or impossible to obtain \cite{Irwin2018}. Citizens are indeed recognized for this key effort \cite{Derrick2022} while favouring new research avenues with open datasets in the context of Computational Social Science \cite{Taylor2016,Leslie2023}. However, new ethical challenges arise \cite{Resnik2015,Taugi2021} and privacy considerations could be overlooked if no specific actions are undertaken. Mobility data must avoid participants identification by for instance inferring home address (see for instance Ref. \cite{Hunter2021,Fan2023,Yang2023}) and geomasking with for instance spatial $k$-anonimity technique is required \cite{Ghinita2010,Wang2020}. In citizen science projects, several aspects are also raised and some measures such as those taken CitSci.org platform could be required as well \cite{Lynn2019}.

In 2012, some of the authors started running citizen science research initiatives to study pedestrian mobility in urban contexts under the name of Beepath experiments. The first initiative tracked participants via a mobile app while wandering through an open space in one of the largest parks in Barcelona (Spain) \cite{Gutierrez2016}. Participants were attending to the local science festival, and several stands sparse within the park were attracting their attention. Data collected allowed us to provide a general modelling framework to analyse reactive and context-dependent factors in a peer-reviewed academic journal \cite{Gutierrez2016}. Several activities were planned to increase public involvement in the research. They included different scientific/technology literacy activities to explain to participants how their data was collected and what was the general purpose of the research. When the participants finished their exploration, personalized reports were shared with each participant, and an aggregated analysis was publicly presented while preserving the participant's privacy.

\begin{figure}
\includegraphics[width=12cm]{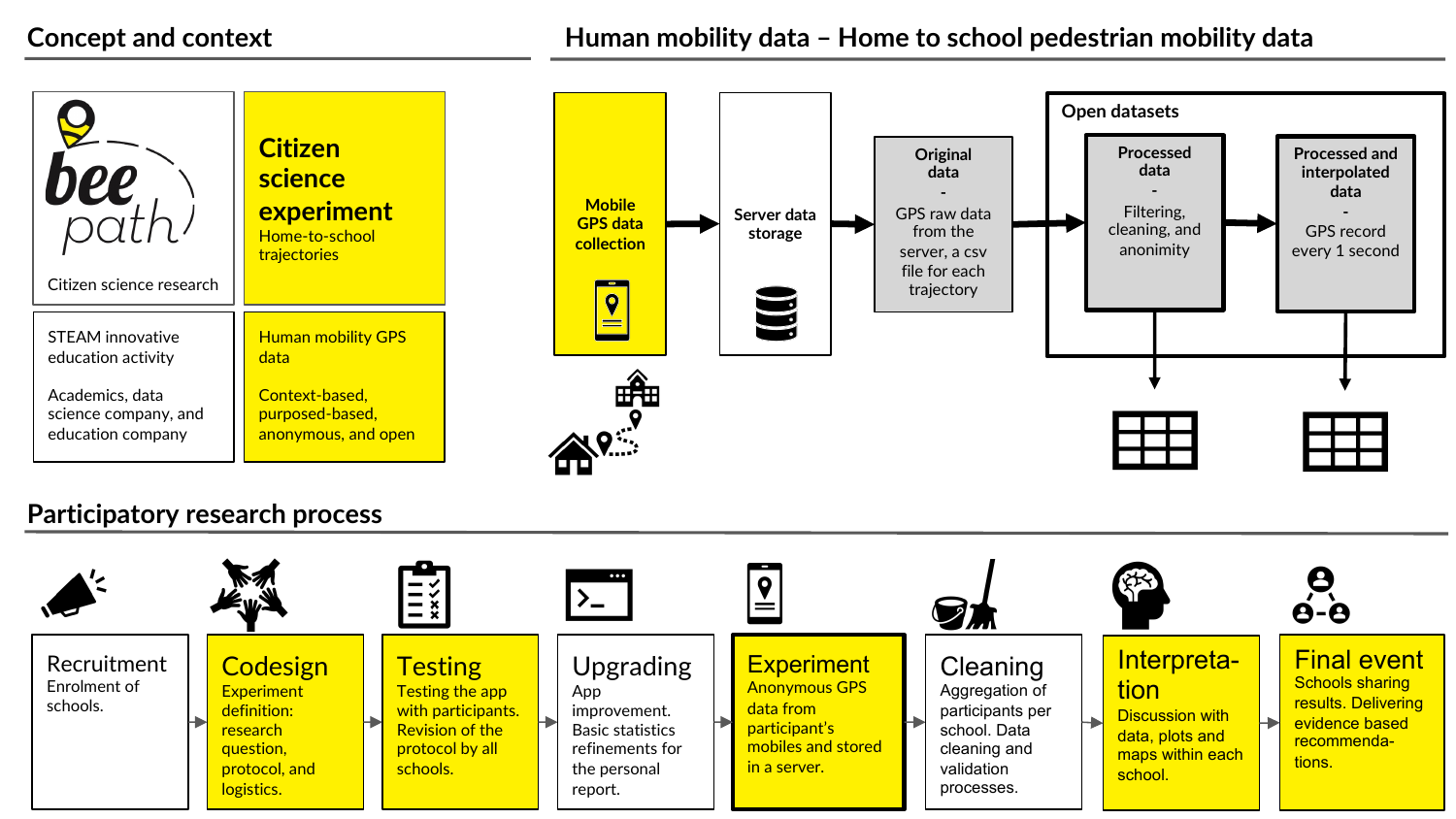}
\caption{{\bf Schematic description of the Beepath citizen science initiative.} We  have combined STEAM education with citizen science situated knowledge co-production to collect GPS data from purposed-based pedestrian mobility. The data files available are reported jointly with the participatory process that has made possible the whole research. The direct participation of students is highlighted in yellow. Grey color highlights the GPS data collected and open data sets are enhanced with a thicker frame.}
\label{fig:summary}
\end{figure}

The Beepath citizen science project developed a new experiment \cite{beepath} under the form of an innovative STEAM activity in formal education (Concept and context in Figure \ref{fig:summary}), as many other citizen science initiatives with schools \cite{Roche2020}. The Beepath experiment we here report wanted to further explore walkability in the ``City of 15-minutes'' \cite{Weng2019}. The decision to work with schools was taken when some schools of the Barcelona Metropolitan area (not the ones participating in the Beepath experiment) were designing their own School Paths (Camins Escolars, in Catalan). These initiatives aimed to encourage children and young people to make their daily journey from home to school on foot or by bicycle, and, most importantly, without adult accompaniment. School Paths were considered a starting point for addressing challenges such as the achievement of a more sustainable mobility model and reflection on children's rights in the city. Schools interested to participate in the Beepath experiment thought that the citizen science research planned could contribute to the ongoing discussion on walkability with GPS data and scientific evidence. The schools broadly hypothesized that their students face several obstacles (e.g., absence of pedestrian crossings, short traffic lights on the green light duration, train railways dividing the neighborhood, or narrow side-walks).

Taking an active learning approach \cite{Land2016,Perello2017,Kloetzer2021}, the Beepath citizen science experiment collected the GPS data with mobile phones in a crowdsourced manner comparable to volunteer geographic information efforts \cite{Haklay2012}. The Beepath citizen science experiment wanted to move beyond the crowd-sourced data collection paradigm and extend participation \cite{Haklay2012} in many other research phases with co-creation and co-design strategies \cite{Senabre2017,Senabre2021}. School participants contributed to the project in different stages (Figure \ref{fig:summary}, in yellow). As reported in the lower part of Figure \ref{fig:summary} (Participatory research process), the co-design phase delivered a protocol and a way to proceed in all research phases. The students became testers of the technology developed and helped run the experiments distributed across ten secondary schools. The students also interpreted the data from their own school group based on their local and situated knowledge and produced data visualizations on their own. They finally presented the results in a public event in front of other school students and municipality representatives. A set of evidence-based recommendations were delivered and merged to reach a wider urban perspective. In Methods, we report these aspects that increase public participation in research and stress the citizen science component that has made it possible to obtain the mobility datasets. Along the Methods descriptions, we also include the actions being taken to preserve privacy of the school  participants. Measures want to keep the location data anonymous and avoid participants to become identifiable as natural persons (avoiding the inference of their home addresses). That is also the reason why we only share processed anonymous GPS data and why we did not collect gender or any other socio-demographic trait.

To exemplify the results from the perspective of the school participants, we briefly report the recommendations delivered by three different schools to municipal representatives (see Ref. \cite{report}, in Catalan). These recommendations were grounded on the investigations done by the students with maps and statistical data features. The first example: The Ferran Tallada public school from Barcelona is in a hilly but still densely populated neighbourhood (El Carmel) and the school building is almost on top of one of the hills. The students emphasized the lack of traffic signals around the school to facilitate pedestrians' safe journeys. Pedestrians followed inefficient movements. The side-walks in this neighbourhood are narrow and the streets are winding with small visibility for car drivers. The students proposed expanding the width of side-walks in well identified locations and the change of traffic direction of particular streets. The second example: Verdaguer school students mostly belong La Ribera neighbourhood. The schools is inside one the largest public parks of Barcelona (Parc de la Ciutadella, 17.42 ha), on the edge of three neighbourhoods. The park has also three entrances and closes at night. Paradoxically, before 8am (when classes start), the only access point to the park (and to the school) was through the more distant door from La Ribera. The data showed that most of the students were making a considerable detour to access to the park as most of the students live in this area. The proposal in this case was simple: open earlier all the doors, specially the one closer to La Ribera. Today, all doors open before 7am. Finally, the third example: the Bellvitge Institute is located at the Bellvitge neighbourhood from L'Hospitalet de Llobregat (the second largest city of the Barcelona metropolitan area). The school is very close to the railways that divide into two parts the neighbourhood. The students do not have many options to cross the railways to reach the school and their trajectories were shown to be highly inefficient. The school jointly with a neighbourhood association sent a formal petition to increase the number of points to cross the railways. The students from this school also participated in an international Science and City workshop within the Barcelona Science and City Biennale (February 7, 2019) to share the experience.

As said, purposed-based pedestrian mobility from quite homogeneous age profile and at a micro-level is very difficult to be obtained. Thanks to citizen science practices, we here report and share unique mobility data from young students' home-to-school trajectories in the Barcelona metropolitan area. This recurrent path is relatively short in time. It shares commonalities with all participants as they have the same destination or origin and belong to a homogeneous age group. As described in the top right corner of Figure \ref{fig:summary} (Human mobility data), GPS data was stored in a server. Datasets were then shared with each participant separately. Datasets were also further processed for scientific research in the way we here report to favour scientific reproducibility and data re-usability (Open datasets in Figure \ref{fig:summary}). In a broader academic scientific perspective, one might suggest a stochastic model which can reliably describe the movement of the participants \cite{Codling2008}. Data can deliver interpretations through appropriate mobility models and contribute to discussions about the most suitable models for pedestrian mobility 
%\cite{models1,models2,models3,models4}. 
\cite{models3,models4,Barbosa2018,Galotti2016}. 
Apart from the velocity we show in Methods, it is possible to further characterize pedestrian mobility with other statistical metrics like reorientation angle or tortuosity \cite{Barbosa2018}. Finally, it can also be of interest to correlate some of the statistical metrics with contextual information such as amount of green or width of the side-walks along similar lines to recent publications \cite{Hunter2021,Fan2023,Yang2023}. The model parameters could be estimated from the empirical data and used to compare mobility in the schools' surroundings in terms of urban structure, more walkable routes (pedestrian streets and green areas) or climate conditions \cite{Melnikov2022}. One starting point for this analysis could be to study whether participants took the most optimal (shortest) route to school (or home) or instead opted for longer routes to choose safer or more walkable streets \cite{Bongiorno2021}.

\section*{Methods}

The Universitat de Barcelona Ethics Committee (IRB00003099) has approved this mobility experiment. All participants read and signed the informed consent and parental/legal guardian consent was also sought as a suitable procedure. No privacy issues have been observed to be in conflict with public release of the underlying processed data.

\subsubsection*{Co-designing the experiment and training}

An open call for participation to schools from the Barcelona metropolitan area was launched. The call resulted in the participation of 427 students (from 14 to 16 years old) and 31 teachers from 3rd and 4th grade of 10 secondary schools, most of them were public schools as reported in Table \ref{tab:schools}. Public school teachers received official recognition from the Consorci d'Educaci\'o de Barcelona, a recognition that counts for promotion to higher professional ranks. 

\begin{table}
\centering
\caption{\label{tab:schools} {\bf List of schools jointly with its code-name (three letters long) and some contextual information.} The school city and district are included in the third and fourth columns. Barcelona has up to 10 districts. Viladecans is a smaller city with around 67,000 inhabitants, and that is the reason why we do not report the district in this case. All cities belong to the Barcelona metropolitan area. ``Ins.'' is an abbreviation of ``Institut'' in Catalan (High School, in English).}
\begin{tabular}{lcll}
\hline\hline
School name & Code-name & City & District\\ \hline
OAK House/Casa del Roure &OAK& Barcelona & Sarri\`a--St Gervasi\\
Ins. Verdaguer & VER& Barcelona& Ciutat Vella\\
Escola Virolai & VIR& Barcelona & Horta--Guinard\'o\\
Col$\cdot$legi Sagrada Fam\'ilia & SAN& Barcelona & Sant Andreu\\
Ins. Pau Claris& IPC& Barcelona & L'Eixample\\
Ins. Bellvitge& BEL& L'Hospitalet & Bellvitge\\
Ins. Montju\"ic& MON& Barcelona& Sants--Monju\"ic\\
Ins. Juan Manuel Zafra&  ZAF& Barcelona & Sant Mart\'i \\
Ins. Ferran Tallada& IFT& Barcelona & Horta--Guinard\'o\\
Col$\cdot$legi Sant Gabriel & SGV& Viladecans & --\\
 \hline
 \hline
\end{tabular}
\end{table}

To start the experiment preparation, two co-design sessions with the involved teachers were performed under the guidance and supervision of education experts (Eduscopi), citizen science researchers (OpenSystems, UB), and data scientists (Dribia). Sessions also introduced citizen science concepts and practices. There was also time devoted to providing basic skills for data visualization analysis for GPS records, and several tools to commit this task were offered. 

The first session with teachers identified shared interests about mobility around schools (3 hours duration). Some logistic aspects related to the mobility experiment were also discussed. Each teacher then transferred to their own students the issues discussed. In-class activities were developed to further discuss, adjust, and validate the joint research to be performed identically by all schools. Some of the questions raised in class were: Which scientific questions could we formulate? What experiment would you like to perform? Initial answers by the students went in the direction about how easy or difficult was to arrive to school from home. Students started to talk about identified obstacles based on their personal perceptions. Through their teachers, the students also had the opportunity to express their own views on the participant's profile, the ideal number of participants, or the best time of the day to perform the experiment. Other topics addressed were related to key statistical measures that can characterize mobility and to the impact that results may have in a neighbourhood level and in a city level. The second session with teachers was performed to put together the perspectives and ideas shared in class (3 hours duration). It was then agreed that the mobility experiment was to be focused on the paths they followed to reach school in a morning or to leave school in an evening, with special attention to pedestrian mobility and walkability.

After the co-design sessions, teachers and students worked further but independently on the execution plan of the mobility study, including the logistics and the calendar for each of the activities planned. Each school had to finalize the design considering the particularities of its own context. They also anticipated in a more detailed manner the impact of the study on their own neighbourhood and how they could communicate the research results locally. 

Doubts and questions were resolved via constant email communication and during a visit to each class group (around 25 students each). The visit was made by one professional researcher from OpenSystems, UB. It allowed us to supervise and validate the approach taken by each school. During the visit, the professional researcher made a presentation on citizen science and pedestrian mobility to all students. The presentation also underlined key technological aspects related to GPS data acquisition and on the mobile app functionalities (see below and Figure \ref{fig:screenshots}). A final debate was organized in terms of data privacy and open science. The scientific protocol (see below) to guarantee scientific rigour and data quality during the experiment was also revised and discussed while preserving the privacy of school participants.

\subsection*{Data acquisition with the mobile phone app}

Each school chose one school day over a two-week period (from November 5 to November 16, 2018). Weather conditions were favourable and allowed us the possibility to compare data among different schools. During the days and hours of the experiment, there were no exceptional climate events (e.g., rain or very low temperatures), and the weather was very similar. To preserve privacy of school participants, the team responsible of running the experiment (some of the authors) did not know exactly which students were collecting their own GPS records. All communication related to the experiment data acquisition was made through the teachers that acted as local coordinators in each school. All schools and school groups involved were large or very large to avoid personal identity inference with the school information. The school participants involved were from two courses (3rd or 4th grade of ESO, Educaci\'on Secundaria Obligatoria, which can be translated as Compulsory Secondary Education) but we did not know which one exactly. The schools had at least two classes per course and each class had 30-35 students. Also, all participants of one school had to perform the experiment the same day. We only allowed to collect data from a pre-established narrow time window of less than an hour. The exact time window was set with teachers. For instance, if participants decided to collect data when going to school, we only opened the server that morning and closed down the server few minutes after the beginning of classes. We activated data collection about 45 minutes before the beginning of classes. 

A total of 262 students finally participated in the experiment with their own mobile phones (see Table \ref{tab:data}). Participation is lower than the number of students involved in the whole research activity (427 students). A relevant number of participating students did not have mobile phones, while others had an old mobile phone which was incompatible with the app version requirements. The schools thus divided research tasks among the students so that, for instance, those who did not have a mobile phone could spend more time analysing data or preparing specific graphics as part of their educational activities related to the project which were also relevant tasks in the joint citizen science research effort. 

\begin{table}
\centering
\caption{\label{tab:data} {\bf Number of participants and GPS data records per each school.} Columns two and three, respectively, report the number of participants and GPS records considering raw data (Original Data). This information is also reported after the filtering and cleaning process in the last three columns (Processed Data). The last column provides GPS interpolated data (GPS int.) which are GPS records reported every second. The code-name (first column) correspondence to each school is provided in Table \ref{tab:schools}. An aggregate with all participants and all GPS records is given in the last row.}
\begin{tabular}{ccccccc}
\hline\hline
 &\multicolumn{2}{c}{Original Data}&&\multicolumn{3}{c}{Processed Data}\\ \cline{2-3} \cline{5-7}
 School&Participants&GPS&&Participants
&GPS&GPS int.\\ \hline
 OAK&$14$&$5,589$&&$1$&$216$&$420$\\
 VIR&$42$&$36,462$&&$2$&$664$&$737$\\
 VER&$11$&$8,870$&&$3$&$1807$&$2,098$\\
 SAN&$15$&$7,980$&&$6$&$1,776$&$2,180$\\
 IPC&$17$&$5,754$&&$7$&$2,187$&$2,502$\\
 BEL&$20$&$16,915$&&$7$&$2,555$&$3,785$\\
 MON&$12$&$4,101$&&$6$&$2,849$&$3,316$\\
 ZAF&$40$&$15,394$&&$13$&$4,081$ &$5,562$\\
 IFT&$31$&$17,859$&&$10$&$5,579$&$5,833$\\
 SGV&$60$&$42,085$&&$28$&$11,508$&$14,620$\\
 \hline
 Total&$262$&$161,009$&&$83$&$33,222$&$41,053$\\
 \hline
 \hline
\end{tabular}
\end{table}

\subsubsection*{Scientific protocol for the experiment}

\begin{figure}
\centering
     \begin{subfigure}[b]{0.3\textwidth}
         \centering
         \includegraphics[width=3cm]{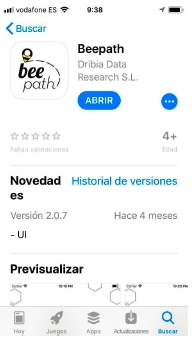}
         \caption{}
         \label{fig:appstore}
     \end{subfigure}
    \begin{subfigure}[b]{0.3\textwidth}
         \centering
         \includegraphics[width=3cm]{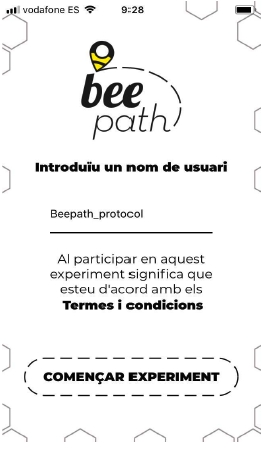}
         \caption{}
         \label{fig:login}
     \end{subfigure}
\begin{subfigure}[b]{0.3\textwidth}
         \centering
         \includegraphics[width=3cm]{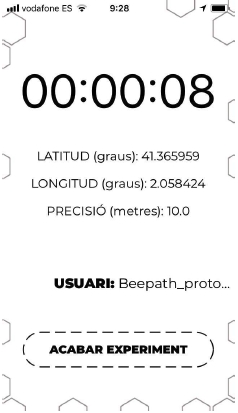}
         \caption{}
         \label{fig:recording}
     \end{subfigure}
\begin{subfigure}[b]{0.3\textwidth}
         \centering
        \includegraphics[width=3cm]{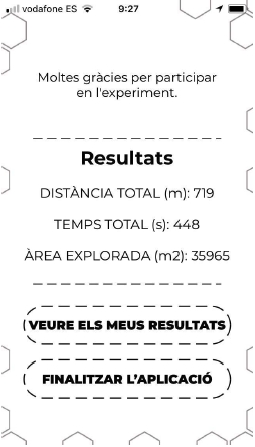}
         \caption{}
         \label{fig:final}
     \end{subfigure}
\begin{subfigure}[b]{0.3\textwidth}
         \centering
        \includegraphics[width=3cm]{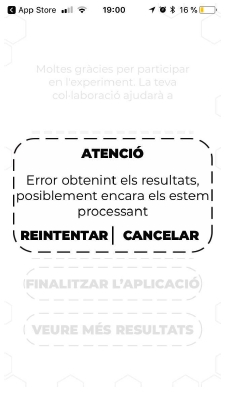}
         \caption{}
         \label{fig:error_1}
     \end{subfigure}
\begin{subfigure}[b]{0.3\textwidth}
         \centering
        \includegraphics[width=3cm]{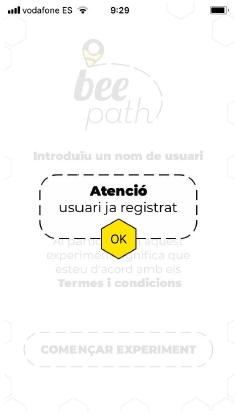}
         \caption{}
         \label{fig:error_2}
     \end{subfigure}
\caption{{\bf Mobile app screenshots for the several steps that the participant had to follow to complete the experiment.} (a) It shows the ``Beepath'' place in the iPhone App Store. (b) It shows the starting screen where participants had to accept terms and conditions with informed consent. The participant also had to insert a nickname (``Usuari'') above the horizontal line and finally click on ``Comen\c car Experiment'' to start the experiment. (c) It shows real-time recording of the trajectory with some information related to GPS. It also gives the option to finish the experiment by clicking on the lowest part of the screen. (d) A summary of basic information about the whole trajectory is provided when the experiment is finished. (e) and (f) A couple of error messages are shown as examples. (e) It shows the case when the participant is still not able to access the data report 
%(see Figure \ref{fig:datareport}) 
because the server is still processing the data. (f) It shows the case when a participant is not allowed to start the experiment with a nickname that has already been used. All texts are in Catalan as the students that participated to the experiment were in Catalan schools.} 
\label{fig:screenshots}
\end{figure}

Each participant used their own mobile phone following a common protocol. All students participating in the experiment by taking the home-to-school trajectory received the following instructions accompanied by some screenshots (see Figure \ref{fig:screenshots}):

{\bf Before the experiment.} (1) Download the app. (2) If you have an iPhone, open the App Store and search for the Beepath app (see a screenshot in Figure \ref{fig:appstore}). If you have an Android, open the Google Play Store and search for the Beepath app. (3) Install the app. (4) Check if the app works properly before the experiment. First, check its functioning on your own. Then repeat the checking with your schoolmates the day that your teacher will announce. (5) In the meantime, report any problem you may have with the app to your teacher.

{\bf During the experiment.} (1) Click on the Beepath app icon when you start the trajectory (see Figure \ref{fig:login}). After reading and accepting the informed consent, you click ``Start the experiment''. The GPS will then start to be recorded. Avoid starting the experiment in indoor spaces as GPS will lose precision. To preserve your privacy, do not start recording data in front of your home. Walk between 5-150 meters before clicking on the app icon to start the experiment. (2) Write your nickname following the established protocol (see the section below and Figure \ref{fig:login}). (3) On the next screen, you should be able to read the latitude in degrees, the longitude in degrees, and the GPS data precision in meters (see Figure \ref{fig:recording}). You will also see the username you wrote in the previous screen. Leave the app on during the trip. The app collects the GPS data from your trip. You do not need to have the Beepath app visible on your screen; you can use other apps during your trip.

{\bf After the experiment.} (1) Press ``End Experiment'' (see Figure \ref{fig:final}). (2) 
The app will then direct you to a web page where you will have a data report: \\ \emph{https://beepath.dribia.com:8080/stats/nickname}. %(see Figure \ref{fig:datareport}). 
(3) To keep the link with you, save the web address in your notebook.

The web address with the personalized data report was unique to each participant. The server was kept active for several months active and allowed the participant to check their own trajectory via a synthesized and automatic data report. Participants could also download their own data. It could be stored in a $csv$ file (with commas) on its own device. Each row included a timestamp (in YYYY:MM:DD HH:MM:SS format) and the GPS records with its latitude and longitude (in degrees). The report provided a map of the tracked GPS data and basic statistical features that included distance covered, time duration, and mean velocity.

Very similar instructions apply to participants performing school-to-home trajectories. The only differences lie in the fact that participants had to click on ``End Experiment'' between 5 to 150 meters before arriving home and to start the experiment inside the school facilities. In both cases and for privacy reasons, data from tests was not stored.

\subsubsection*{Nicknames and school code names}

As mentioned in the protocol, each participant received a number assigned by each teacher. Nicknames were used to preserve the privacy and anonymity of the participants. They were asked to add a nickname that contained this number and additional key information: the mean of transport (bus, walking, car, scooter, bike, metro, train...), whether they made the trip alone or together with other schoolmates, and the school code-name. An example of what a participant had to include as a nickname when starting the experiment could read in the following manner: {\tt ZAF\_0001\_WALK\_ALONE}, which corresponds to Juan Manuel Zafra school participant number 1 (assigned by the teacher) that walked to school alone. As scientists, we could not know the participants' identities. The information characterizing each trajectory finally appears in the $csv$ filename. It was encoded jointly with the date of the trip. An example could read as follows: {\tt 2018-11-05\_sgv\_2603.csv}.

School codenames are: Institut Juan Manuel Zafra (ZAF), Col$\cdot$legi Sagrada Fam\'ilia Sant Andreu (SAN), Institut Montju\"ic (MON), Institut Verdaguer (VER), Institut Ferran Tallada (IFT),  OAK House/Casa del Roure (OAK), Escola Virolai (VIR), Institut Pau Claris (IPC), Col$\cdot$legi Sant Gabriel de Viladecans (SGV), Institut Bellvitge (BEL). They are also reported in Table \ref{tab:schools}.

\subsubsection*{Supervision, monitoring and support}

Before the experiments, data scientists from Dribia monitored and supervised the testing made by each school with some basic statistical analysis with aggregated data and by immediately erasing the data collected. There were some errors reported by the participants themselves but always throughout their teachers. They also implemented modifications in the app and in the data personalized report based on students' and teachers' feedback. During the experiment, data scientists from Dribia also monitored and supervised the progress of the data recording in real-time and did not collect any data out of the pre-established time window and which was not following the pre-established nick naming structure. A server was specifically prepared for the experiments, and the CPU capacity was increased during the weeks of the experiments. 

After each experiment, making a first revision of all trajectories and some basic cleaning procedures, data scientists from Dribia aggregated the data per school. Every teacher thus received anonymous $csv$ files with all trajectories from their school. The students took data and reflected on it. Maps and other visualizations were created in a standardized info-graphics format designed by Eduscopi. The template gave space to show plots, explain the results, and deliver recommendations related to pedestrian mobility in each neighborhood \cite{report}. All data manipulation was mediated by teachers and professional scientists did not have any direct contact with school participants in the data interpretation effort made by each school.

\subsubsection*{Actionable data to deliver policy recommendations}

%\subsubsection*{\textcolor{blue}{Actionable knowledge}}

At the BarcelonActiva Auditorium, a 2-hours duration closing event was held with about 200 students and teachers in the audience. Student representatives for each school showed their own results and shared their own conclusions. A joint report collecting the info-graphics ($pdf$ file in Catalan, shared jointly with the datasets) \cite{report} and the set of policy recommendations were delivered to the Technical Director of Digital Innovation at the City Council of Barcelona and the institute of Barcelona digital city (i.lab). Some of the schools also used the analysis to go to public municipal or district authorities and ask for some very specific actions (generally at an urban micro-level). Some of the results are reported in the previous Background \& Summary section.

\subsection*{Filtering, cleaning, and interpolating GPS data}

As scientists, we are interested in purposed-based pedestrian mobility, but not all participants followed this type of mobility. Furthermore, in some cases, the data acquisition process appeared incomplete or partially failed for technical reasons. It was thus necessary to make a thorough filtering process. Also, for scientific purposes, it is also very much convenient to have GPS data with periodic timestamps via some interpolation.

Let us, however first define the following key variables that characterize the mobility of each participant. Distance between consecutive GPS records reads
\begin{equation}
d(t)=|\vec{r}(t+\Delta(t))-\vec{r}(t)|, 
\label{d}
\end{equation}
where we have an irregular duration of consecutive GPS timestamps $\Delta(t)$ and $\vec{r}(t)$ is the GPS two-coordinate vector of a pedestrian position at time $t$ (a given GPS time stamp). We thus define instantaneous velocity as
\begin{equation}
v(t)=\frac{|\vec{r}(t+\Delta(t))-\vec{r}(t)|}{\Delta(t)}=\frac{d(t)}{\Delta(t)},
\label{v}
\end{equation}
and total distance covered by one participant as
\begin{equation}
D=\sum_{\{t\}} |\vec{r}(t+\Delta(t))-\vec{r}(t)|= \sum_{\{t\}} d(t),
\label{D}
\end{equation}
where we sum over all timestamps ${\{t\}}$ from a trajectory except from the last one. Finally, the computing of the trajectory duration thus reads
\begin{equation}
T=\sum_{\{t\}} \Delta(t).
\label{T}
\end{equation}

The filtering and cleaning process is divided into five successive steps. First, non-pedestrian users are filtered out by checking the means of transport they use. Second, the GPS paths are displayed on maps, and the invalid ones are removed (those that do not form a well-defined origin-destination trajectory). Third, the remaining trajectories are cleaned of noisy GPS locations at the beginning and/or end of the path due mostly to GPS connection problems (when they are still located at schools). This is done by visualizing the routes on maps along with the study of the distances (cf. Eq. (\ref{d})) and velocities (cf. Eq. (\ref{v})), which typically present outliers at these noisy locations. Finally, again to preserve participants privacy, we mask the location data (GPS records) by removing a random amount of records (the first 20-50 seconds of the trajectory) of the movement in home-to-school trajectories and (the last 20-50 seconds of the trajectory) of the movement in school-to-home trajectories. The spatial $k$-anonymity technique is used to measure the disclosure risk \cite{Ghinita2010}. This privacy-by-design measure is taken as an addition to the protocol instruction described above to start the app recording about 5-150 meters  after leaving (or before reaching) home. The fifth step consists in interpolating the GPS locations linearly to have all records uniformly separated by one second.

Initial data reported in Table \ref{tab:data} is reduced to $83$ participants and $33,222$ GPS locations ($41,053$ after interpolation). The details of the process are reported in the forthcoming sections.

\subsubsection*{Non-pedestrians removal}

The nickname (as described in the previous sections) not only makes sure that data is anonymous, but also includes trip details such as the school code name and the type of transport being used. Since we are exclusively interested in purposed-based pedestrian mobility, we removed from the data set those participants who did not follow this type of mobility. For instance, the participant {\tt 2018-11-05\_sgv\_0601\_tren} used the train as a transport mode. Additional attention is required for those participants that did not indicate the means of transport. Through statistics, we can infer that some of these participants used another means of transport rather than walking. We thus remove abnormal averaged velocities.

We finally filter out a total of $105$ non-pedestrian trajectories corresponding to a total number of $83,009$ GPS locations.

\subsubsection*{Invalid trajectories and non-origin-destination trajectories}

\begin{figure}[t] 
     \centering
    \begin{subfigure}[b]{0.42\textwidth}
         \centering
         \includegraphics[width=\textwidth]{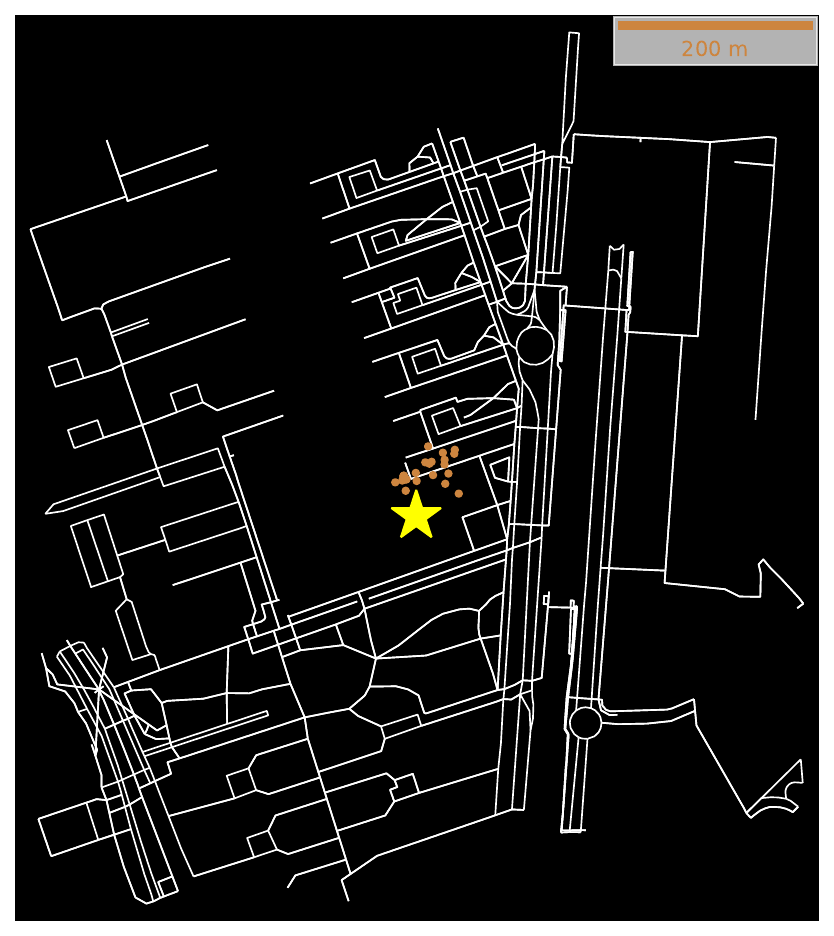}
         \caption{}
         \label{fig:invalid_journey_bellvitge}
     \end{subfigure}
     \begin{subfigure}[b]{0.47\textwidth}
         \centering
         \includegraphics[width=\textwidth]{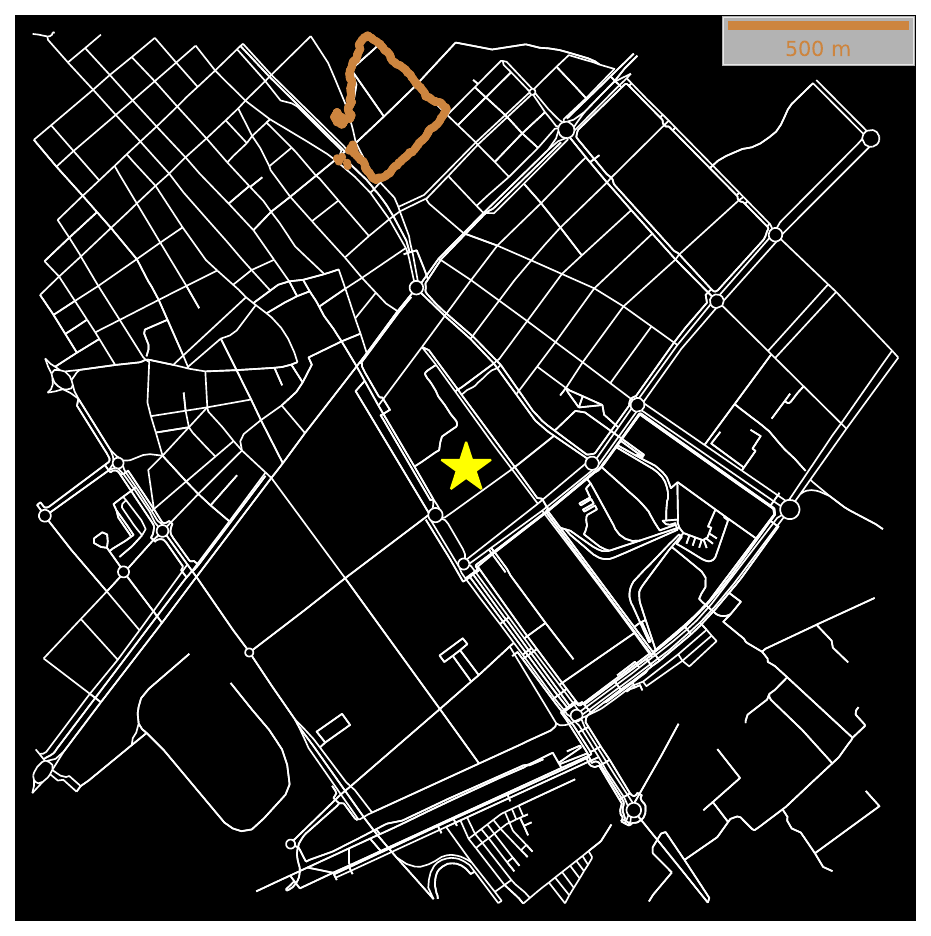}
         \caption{}
         \label{fig:invalid_journey_sgv}
     \end{subfigure}
     \caption{{\bf Example of an invalid trajectory.} (\textbf{a}) Map visualization of an invalid trajectory from a BEL participant, with only a few GPS locations scattered around the school (yellow star, school). (\textbf{b}) Map visualization of a trajectory with no-origin destination from home to school, from a participant of SGV.}
     \label{fig:invalid_journey}
\end{figure}

We visually explore each trajectory record separately, displaying the GPS locations on maps. Several trajectory records do not follow a clear path, possibly due to problems with the app or with the GPS satellite connection. Invalid trajectories are also due to human error when recording the route (see Figure \ref{fig:invalid_journey_bellvitge}). These records are removed jointly with those that neither start nor end in one of the schools (see Figure \ref{fig:invalid_journey_sgv}). 

This second step has removed a total amount of $7,500$ GPS locations, corresponding to $43$ participants.

\subsubsection*{Outliers}

\begin{figure}[t]
     \centering
     \begin{subfigure}[b]{0.40\textwidth}
         \centering
         \includegraphics[width=\textwidth]{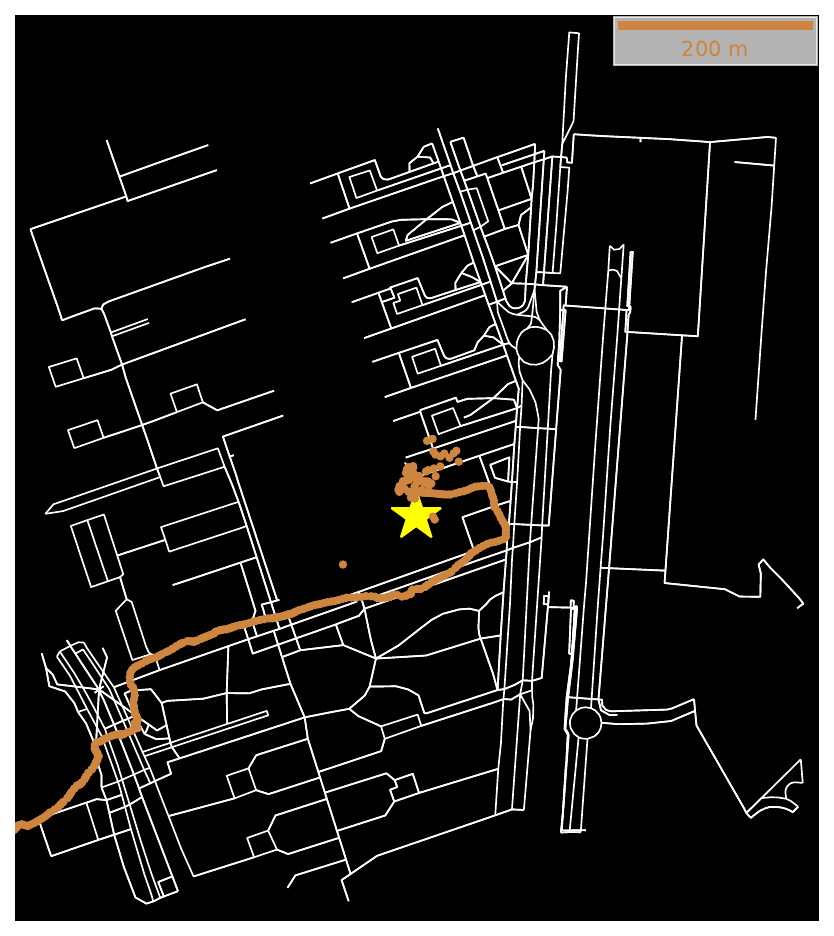}
         \caption{}
         \label{fig:outlier_bel1}
     \end{subfigure}
     \begin{subfigure}[b]{0.58\textwidth}
         \centering
         \includegraphics[width=\textwidth]{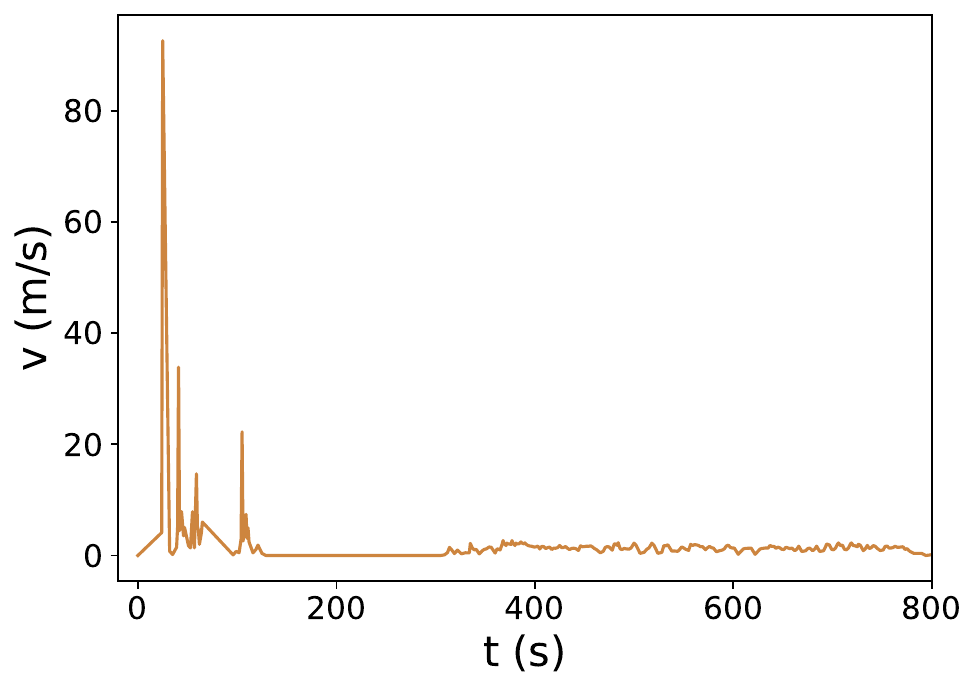}
         \caption{}
         \label{fig:outlier_bel2}
     \end{subfigure}
     \begin{subfigure}[b]{0.40\textwidth}
         \centering
         \includegraphics[width=\textwidth]{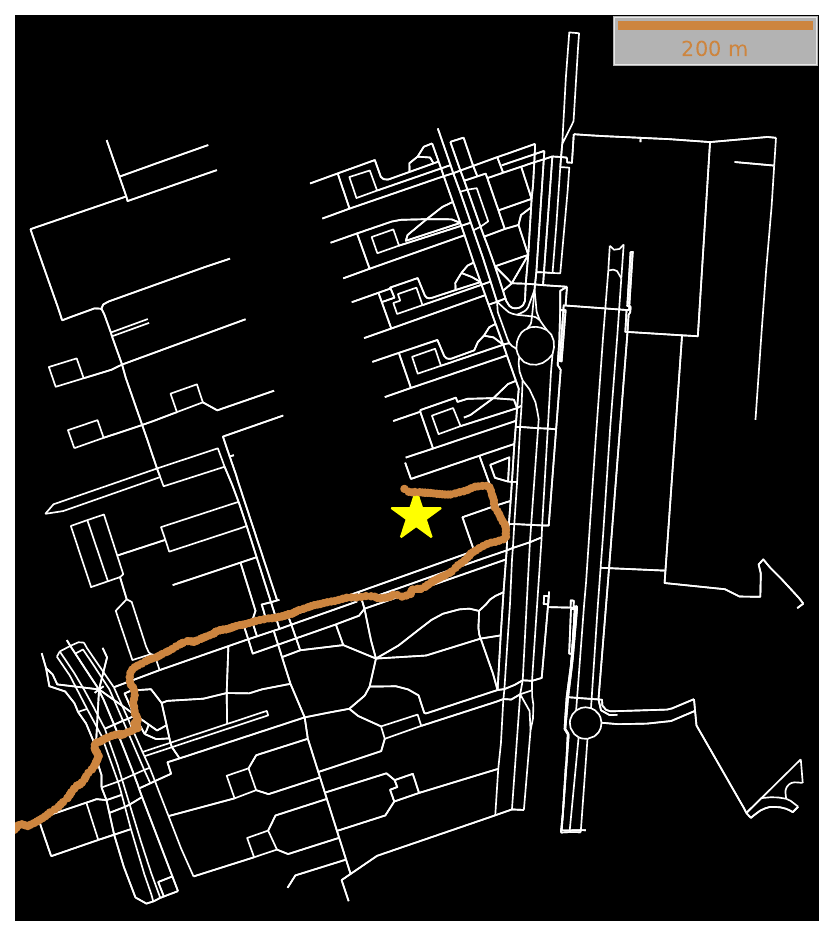}
         \caption{}
         \label{fig:outlier_bel3}
     \end{subfigure}
     \begin{subfigure}[b]{0.58\textwidth}
         \centering
         \includegraphics[width=\textwidth]{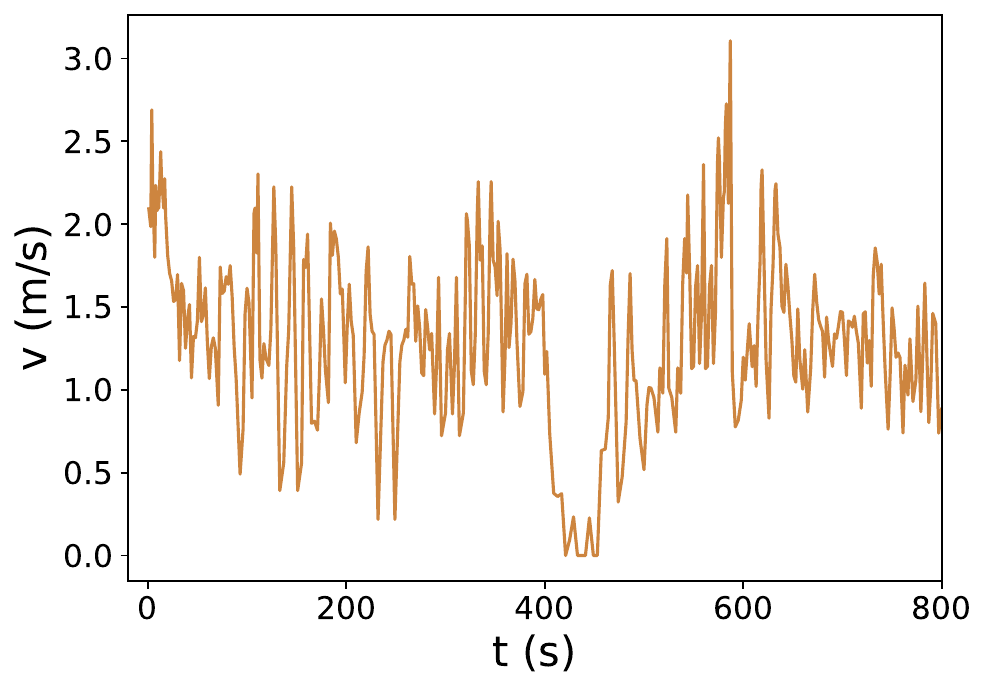}
         \caption{}
         \label{fig:outlier_bel4}
     \end{subfigure}
     \caption{ {\bf Example of outlier clean-up.} (\textbf{a}) Trajectory of a participant from BEL with outliers at the beginning of the route (yellow star, school). (\textbf{b}) Time series of the participant's instantaneous velocity with outliers. (\textbf{c}) Visualization of the trajectory and (\textbf{d}) Temporal series of the instantaneous velocity of the same participant after the outlier clean-up process.}
     \label{fig:outlier_bel}
\end{figure}

When computing time increments $\Delta(t)$ between consecutive GPS timestamps, GPS consecutive distances $d(t)$ (cf. Eq. (\ref{d})) and instantaneous velocities $v(t)$ (cf. Eq. (\ref{v})), it is also possible to detect large outliers. These outliers are frequent in first GPS records as the GPS activation generally provides some initial noisy records. This is evident when displaying the trajectory of one participant on a map as shown in Figure \ref{fig:outlier_bel1}. Figure \ref{fig:outlier_bel2} thus shows its extremely large velocity values, exceeding even $90$ m/s. Other outliers are also sometimes detected very close to the end of the trajectory due to the fact that the participant is inside the school building, where GPS has much less precision or where the mobile is automatically connected to WiFi. Then GPS locations are accumulated in the same area for a while. 

All these outliers are carefully analyzed and removed (see Figure \ref{fig:outlier_bel3} and Figure \ref{fig:outlier_bel4} as an example). Still, there are certain individual trajectories that contain too many outliers, thus making it impossible to create a meaningful origin-destination trajectory once the outliers are removed. In these cases, we discard the whole trajectory.

In total, $31$ complete journeys and $34,409$ GPS outliers are being removed. Therefore, we ended up with $83$ individual trajectories and a total amount of $36,091$ records.

\subsubsection*{Geomasking and spatial $k$-anonymity}
\begin{figure}
\includegraphics[width=11cm]{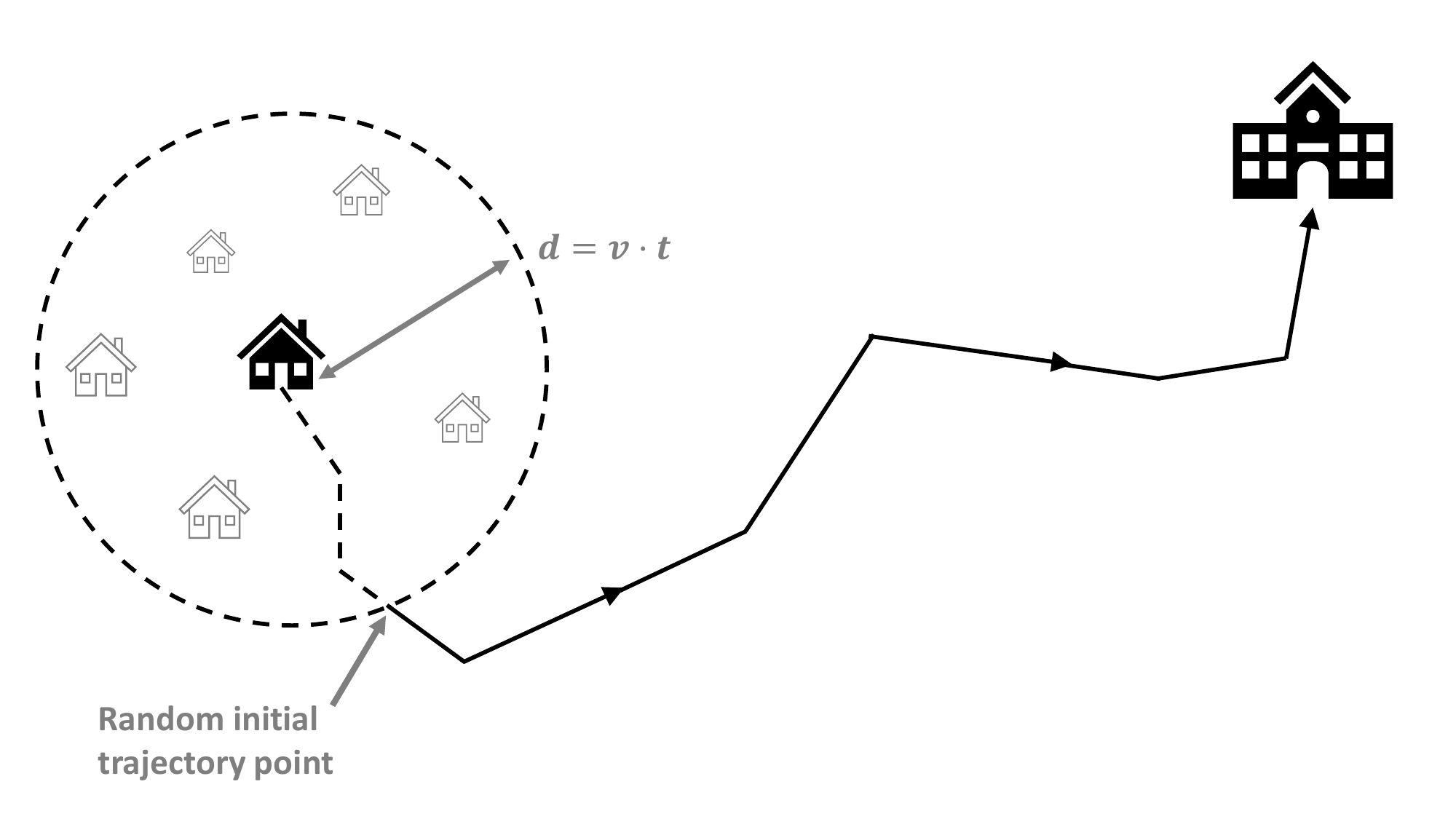}
\caption{{\bf Schematic $k$-anonymity technique for geomasking.} We consider a circle of radius $d=v\cdot t$ around the initial location of home-to-school trajectories, assuming $v=1.5m/s$ and $t$ a random quantity between $20$ and $50$ s. The first $t$ locations (in seconds) of movement are then removed from the trajectory. The new initial trajectory point is such as the probability of revealing the original location (home) is  small (we have considered an order of magnitude of $10^{-3}$). The same procedure applies for school-to-home trajectories, but considering the last $t$ seconds (locations) of the trajectory.}
\label{fig:k-anonymity}
\end{figure}

Location data under the form of GPS data can potentially reveal personal identity (through home address, in our case). We have no mean to make sure that participants have followed the protocol instruction described above to start the app recording about 5-150 meters from their home (or to finalize the GPS recording 5-150 meters before reaching their home). In the remaining cleaned trajectories, we thus perform a further filtering to keep the anonymity of the participants. We have used the spatial $k$-anonymity technique \cite{Ghinita2010,Wang2020} to mask the location data within a certain urban area. The technique strips away GPS records that could identify home address as shown in Fig. \ref{fig:k-anonymity}. $K$-anonymity is thus here characterized as the number of home units $k$ within a given area. Therefore, $1/k$ quantifies the disclosure risk, that is roughly the probability of revealing the correct location of the participant's home.

Table \ref{tab:disclosure_risk} provides the urban surface and the number of housing units of a given district (or city) obtained from different reports accessible online \cite{statistics_bcn, statistics_viladecans, statistics_hospi}. With this information, we can estimate the density of housing units as shown in the third column in Table \ref{tab:disclosure_risk}. Then, last column approximately obtains the number $k$ of homes within a given circle of radius $d=v\cdot t$. To make it simple and as we only want a rough estimation, we can take constant velocity $v$, identical for all participants. Then, based on the disclosure risk $1/k$ willing to assume, we can finally decide the time period $t$ in seconds we strip away from indidvidual GPS records at the beginning of the home-to-school trajectories (or at the end for school-to-home trajectories, but always when participants have non-zero velocity). If we approximately take $v=1.5$ m/s for all participants (see Data Records Section to check that this is a reasonable choice) and use $t=50$ s, we can obtain an average disclosure risk $\langle 1/k \rangle=2.72\times 10^{-3}$, which is a comparable order of magnitude taken by other publications using GPS data \cite{Hunter2021,Fan2023,Yang2023}. Table \ref{tab:disclosure_risk} shows the details for each of the districts and cities. We have therefore chosen a random value of $t$ between $20$ seconds and $50$ seconds for each participant (each trajectory). Table \ref{tab:disclosure_risk2} reports the number of GPS records removed with the spatial $k$-anonymity and how the technique shortens the total duration of the trajectories.

In total, $2,869$ GPS records are being removed. Therefore, we ended up with $83$ individual trajectories and $33,222$ GPS records data after the implementation of the $k$-anonimity technique.

\begin{table}[t]
\centering
\caption{\label{tab:disclosure_risk} {\bf Disclosure risk for each district of Barcelona or city.} Columns respectively report the district of Barcelona or city where the school is located, the urban surface in km$^2$ units, the number of housing units, the density of housing in $1/\mbox{km}^2$, and the $1/k$ disclosure risk, which is calculated multiplying the density of housing units by the surface area of a circle of radius $d=v\cdot t$, with $v=1.5$ m/s and $t=50$ s. The averaged disclosure risk is $2.72\times 10^{-3}$. L'Hospitalet and Viladecans are cities, not districts of Barcelona.}
\begin{adjustbox}{width=\textwidth}
\begin{tabular}{lcccc}
\hline\hline
District or city & Urban surface ($\mbox{km}^{2}$) & Housing units & Housing units density ($1/\mbox{km}^{2}$) & $1/k$ \\ \hline
Sarri\`a & $6.105$ &  $74,729$ & $12,240.62$ & $4.62\times 10^{-3}$ \\
Ciutat Vella & $1.296$ & $55,663$ & $42,949.85$ & $1.32\times 10^{-3}$ \\
Horta-Guinard\'o & $2.898$ & $78,367$ & $27,041.75$ & $2.09\times 10^{-3}$ \\
Sant Andreu & $1.906$ & $70,056$ & $36,755.51$ & $1.54\times 10^{-3}$\\
L'Eixample & $3.719$ & $143,403$ & $38,559.55$ & $1.47\times 10^{-3}$ \\
L'Hospitalet & $3.010$ & $111,363$ & $36,997.67$ & $1.53\times 10^{-3}$ \\
Sants-Montju\"{i}c & $2.516$ & $90,449$ & $35,949.52$ & $1.57\times 10^{-3}$ \\
Sant Mart\'i & $2.889$ & $111,536$ & $38,607.13$ & $1.47\times 10^{-3}$ \\
Viladecans & $3.811$ & $24,221$ & $6,355.95$ & $8.90\times 10^{-3}$ \\
\hline
\hline
\end{tabular}
\end{adjustbox}
\end{table}

\begin{table}[t]
\centering
\caption{\label{tab:disclosure_risk2} {\bf Number of GPS records and average trip duration for each school after spatial $k$-anonymity.} Columns respectively report the district of Barcelona or city, the school code-name, the number of participants, the number of GPS records and the average trip duration in seconds, before and after applying the spatial $k$-anonymity. The number of GPS records after applying the spatial $k$-anonymity also includes the $\%$ reduction. L'Hospitalet del Llobregat and Viladecans are cities, not districts of Barcelona.}
\begin{adjustbox}{width=\textwidth}
\begin{tabular}{lcccccc}
\hline\hline
District or city & School & Participants & GPS & $k$-GPS & $\langle T \rangle$ (s) & $k$-$\langle T \rangle$ (s) \\ \hline
Sarri\`a & OAK & $1$ &  $246$ & $216$ ($12\%$)& $474$ & $419$ \\
Ciutat Vella & VER & $3$ & $1,926$ & $1,807$ ($6\%$) & $741$ & $698$ \\
Horta-Guinard\'o & VIR & $2$ & $721$ & $664$ ($8\%$)& $403$ & $367$ \\
Sant Andreu & SAN & $6$ & $1,968$ & $1,776$ ($10\%$) & $402$ & $362$ \\
L'Eixample & IPC & $7$ & $2,460$ & $2,187$ ($11\%$)& $398$ & $356$ \\
L'Hospitalet & BEL & $7$ & $2,782$ & $2,555$ ($8\%$) & $590$ & $539$ \\
Sants-Montju\"{i}c & MON & $6$ & $3,114$ & $2,849$ ($9\%$) & $607$ & $552$ \\
Sant Mart\'i & ZAF & $13$ & $4,485$ & $4,081$ ($9\%$) & $469$ & $427$ \\
Horta-Guinard\'o & IFT & $10$ & $5,907$ & $5,579$ ($6\%$) & $616$ & $582$ \\
Viladecans & SGV & $28$ & $12,482$ & $11,508$ ($8\%$) & $565$ & $521$ \\ \hline
Total & & $83$ & $36,091$ & $33,222$ ($8\%$) & $537$ & $494$ \\
 \hline
 \hline
\end{tabular}
\end{adjustbox}
\end{table}

\subsubsection*{Temporal gaps and linear interpolation}

The mobile app is designed to collect data uniformly every second ($\Delta(t)=1$ s). However, on some occasions, the trajectory data sets contain larger time intervals between consecutive GPS records. This can be attributed to errors in the GPS connection, having subsequent antenna signal jumps. They can also be attributed to mobile app pauses. $17\%$ of the data is separated by more than 1 second, but $99.8\%$ of the GPS records are separated by $\Delta (t) \leq 4 s$. Large temporal gaps are, therefore, statistically irrelevant. 

Even if temporal gaps are small, scientific analysis on pedestrian micro-mobility may require to have constant periodicity in consecutive GPS records. This can be the case in a study on instantaneous velocity. To have a record of every second, we perform a linear interpolation. The number of GPS locations is then increased from $33,222$ to $41,053$ records. This procedure does not reveal any remarkable effects on the dynamics as shown in Figure \ref{fig:interpolated_figures}. Table \ref{tab:data} shows the number of participants and GPS locations for each school after processing the data and after the linear interpolation procedure.

Tables \ref{tab:data2} and \ref{tab:data3} include the main statistical indicators for each school: the distance travelled, the time spent, and the velocity (after processing the data but before linear interpolation).

\begin{table}[t]
\centering
\caption{\label{tab:data2} {\bf Distance and duration of the trajectories for each school}. Mean distance traveled (cf. Eq. (\ref{D})), the shortest and the largest trajectory, the mean amount of time spent (cf. Eq. (\ref{T})), the quickest and the fastest trajectory for each school and after filtering and processing the data (before linear interpolation).}
\begin{adjustbox}{width=\textwidth}
\begin{tabular}{ccccccc}
\hline\hline
School & $\langle D \rangle$ (m) & $D_{min}$ (m) & $D_{max}$ (m) & $\langle T \rangle$ (s) & $T_{min}$ (s) & $T_{max}$ (s) \\ \hline
 OAK & $639$ & $639$ & $639$ & $419$ & $419$ & $419$  \\
 VER & $1105 \pm 194$ & $694$ & $1515$ & $698 \pm 143$ & $462$ & $1040$  \\
 VIR & $391 \pm 67$ & $297$ & $485$ & $368\pm 46$ & $302$ & $433$  \\
 SAN & $541 \pm 103$ & $168$ & $945$ & $362\pm 61$ & $172$ & $559$  \\
 IPC & $564 \pm 89$ & $291$ & $989$ & $356\pm 55$ & $171$ & $589$ \\
 BEL & $679 \pm 83$ & $279$ & $927$ & $540\pm 62$ & $250$ & $775$\\
 MON & $700 \pm 79$ & $367$ & $1007$ & $552\pm 61$ & $300$ & $753$\\
 ZAF & $607 \pm 92$ & $178$ & $1439$ & $427\pm 89$ & $107$ & $1228$ \\
 IFT & $824 \pm 188$ & $84$ & $2305$ & $582\pm 126$ & $67$ & $1582$ \\
 SGV & $756 \pm 64$ & $195$ & $1788$ & $521\pm 54$ & $126$ & $1571$\\
 \hline
 \hline
\end{tabular}
\end{adjustbox}
\end{table}

\begin{table}[t]
\caption{\label{tab:data3} {\bf Instantaneous velocity for each school}. Mean, the smallest and largest value of the instantaneous velocity (cf. Eq. (\ref{v})) for each school and after filtering and processing the data (before linear interpolation).}
\centering
\begin{tabular}{cccc}
\hline\hline
School & $\langle v \rangle$ (m/s) & $v_{min}$ (m/s) & $v_{max}$ (m/s)\\ \hline
 OAK & $1.57\pm 0.02$  & $0.45$ & $2.42$ \\
 VER & $1.64\pm 0.02$ & $0.02$ & $5.26$  \\
 VIR & $1.12\pm 0.02$  & $0.06$ & $3.05$ \\
 SAN & $1.61\pm 0.02$ & $0.05$ & $7.57$  \\
 IPC & $1.64\pm 0.02$ & $0.07$ & $7.77$  \\
 BEL & $1.44\pm 0.02$ & $0.02$ & $7.69$  \\
 MON &  $1.33\pm 0.01$ & $0.07$ & $3.65$ \\
 ZAF & $1.64\pm 0.02$ & $0.02$ & $8.57$ \\
 IFT & $1.45\pm 0.01$ & $0.07$ & $7.88$  \\
 SGV & $1.54\pm 0.01$ & $0.01$ & $8.15$ \\
 \hline
 \hline
\end{tabular}
\end{table}

\subsection*{Code availability}

The \url{https://github.com/ferranlarroyaub/Beepath-Schools} repository holds the Python code and scripts to process the input data \cite{dataverse} and to replicate the statistical analysis and the figures. 
The $3.8$ Python version is used to build the code with the main libraries: {\tt networkx} and {\tt osmnx} to plot the trajectories on OpenStreet maps. {\tt Pandas} and {\tt numpy} to process, clean, and analyze the data in Data-Frame format and perform the basic statistic calculations. {\tt Scipy} for more advanced calculations such as fitting models to the empirical data and {\tt matplotlib} for plotting purposes. The Python code is built in different Jupyter notebook files which contain a detailed description of the study and the code documentation.

\section*{Data Records}
The data repository \cite{dataverse} contains the collected and processed datasets, distributed in two different folders. 

%\begin{table}[t]
%\caption{\label{tab:example_file} {\bf Example of a raw \textit{csv} file display.} This is an SGV pedestrian participant with 604 records (GPS locations). }
%\begin{adjustbox}{width=\textwidth}
%\begin{tabular}{cccccccc}
%\hline\hline
%& course & haccuracy & latitude & longitude & speed & time & nickname %\\ \hline
%0 & $20.7$ & $25.0$ & $41.319088$ & $2.021116$ & $1.49$ & 2018-11-09 19:22:02 & sgv\_0802\_peu \\
%1 & $63.1$ & $19.0$ & $41.319070$ & $2.021156$ & $1.01$ & 2018-11-09 19:22:04 & sgv\_0802\_peu  \\
%2 & $83.1$ & $16.9$ & $41.319073$ & $2.021162$ & $1.07$ & 2018-11-09 19:22:05 & sgv\_0802\_peu \\
%3 & $0.0$ & $15.0$ & $41.319071$ & $2.021171$ & $0.0$ & 2018-11-09 19:22:06 & sgv\_0802\_peu \\
%... & ... & ... & ... & ... & ... & ... & ... \\
%601 & $338.6$ & $9.0$ & $41.320196$ & $2.020269$ & $0.19$ & 2018-11-09 19:33:51 & sgv\_0802\_peu \\
%602 & $328.7$ & $7.0$ & $41.320192$ & $2.020261$ & $0.24$ & 2018-11-09 19:33:52 & sgv\_0802\_peu \\
%603 & $330.9$ & $6.0$ & $41.320172$ & $2.020267$ & $0.15$ & 2018-11-09 19:33:53 & sgv\_0802\_peu\\
% \hline
% \hline
%\end{tabular}
%\end{adjustbox}
%\end{table}

The {\tt processed data} folder contains the 83 $csv$ files reporting the participant trajectories after the filtering and cleaning process (i.e., removing non-pedestrian participants, removing invalid and non-origin-destination trajectories, cleaning outliers and removing the first or last 20-50 seconds of movement). Processed files are saved with the suffix {\tt processed} (e.g.: {\tt 2018-11-05\_sgv\_0802 \_peu\_processed.csv}). Each of the $csv$ files has 10 columns, 7 of which correspond to the GPS raw data collected in the experiment by the Android and IOS platform server (see Table \ref{tab:columns_raw_data}). The other three columns correspond to the time difference between GPS timestamps, the distance between consecutive GPS locations and the corresponding instantaneous velocity (see Table \ref{tab:columns_processed_data}). We remark that the calculation of $\Delta(t)$, $d(t)$ and $v(t)$ is done in a time-advanced way (cf. Eqs. (\ref{d}) and (\ref{v})): the time difference of the record at the location {\tt i} is the time difference between the timestamps at locations {\tt i+1} and {\tt i} (the same applies for the distance and velocity). For this reason, the last record (last row) does not have these three values reported. Table \ref{tab:example_file2} shows an example of processed data table.

\begin{table}[t]
\caption{\label{tab:columns_raw_data} {\bf Columns of the raw \textit{csv} files.} Description of the 7 columns of the $csv$ files from the original data, collected by the Android/IOS app servers.}
\begin{adjustbox}{width=\textwidth}
\begin{tabular}{ll}
\hline\hline
Column & Description \\ \hline
course &  The direction in which the device is traveling, measured in degrees and relative to due north.\\ 
haccuracy & The radius of uncertainty for the location, measured in meters. \\ 
latitude & Latitude coordinate of each record in degrees. \\ 
longitude & Longitude coordinate of each record in degrees. \\ 
speed & Instantaneous velocity of the device, in meters per second.\\ 
time & Timestamp of each record in YYYY:MM:DD HH:MM:SS format.  \\ 
nickname & Anonymous nickname of the participant. \\
 \hline
 \hline
\end{tabular}
\end{adjustbox}
\end{table}

For convenience and data quality testing, we are interested in having the data processed with and without linear interpolation separately. The folder {\tt interpolated data} contains the same $csv$ processed files but with linear interpolation. The suffix {\tt interpolated} is added to each filename instead of {\tt processed} (e.g.: {\tt 2018-11-05\_sgv\_0802\_peu\_interpolated.csv}). After completing the procedure of linear interpolation, all the records are equally spaced. They are periodic, every 1 second. Therefore, in the processed and interpolated $csv$ files, the column $\Delta (t)$ is always $1.0$, and the columns {\tt d} (now distance covered in 1 second) and {\tt v} (velocity) provides the same value.

\begin{table}[t]
\caption{\label{tab:columns_processed_data} {\bf New columns of the filtered and processed \textit{csv} files.} Description of the $3$ new columns. }
\begin{adjustbox}{width=\textwidth}
\begin{tabular}{ll}
\hline\hline
Column & Description \\ \hline
$\Delta t$ & Time difference between consecutive timestamps $t$, in seconds. \\ 
$d$ &  Distance at time $t$ between consecutive GPS locations, in meters.\\
$v$ & Instantaneous velocity at time $t$ (distance over time-lapse), in meters/second.  \\ 
 \hline
 \hline
\end{tabular}
\end{adjustbox}
\end{table}

\begin{table}[t]
\caption{\label{tab:example_file2} {\bf Example of a processed and cleaned \textit{csv} file display.} This is a ZAF pedestrian participant, with $220$ records. }
\begin{adjustbox}{width=\textwidth}
\begin{tabular}{ccccccccccc}
\hline\hline
& course & haccuracy & latitude & longitude & speed & time & nickname & $\Delta t$ & $d$ & $v$ \\ \hline
0 & $256.221925$&	$12.0$&	$41.413296$&	$2.187355$&	$0.775147$ & 2018-11-09 09:53:25 & zaf\_0067\_peu & $1.0$ & $2.367819$ & $2.367819$\\
1 & $273.831081$&	$25.0$&	$41.413310$&	$2.187333$&	$0.941530$ & 2018-11-09 09:53:26 & zaf\_0067\_peu & $1.0$ & $2.149949$ & $2.149949$ \\
2 & $275.834512$&	$25.0$&	$41.413319$&	$2.187310$&	$0.941530$ & 2018-11-09 09:53:27 & zaf\_0067\_peu & $1.0$ & $3.443628$ & $3.443628$ \\
3 & $253.445671$& 	$25.0$	& $41.413342$& 	$2.187283$	& $0.805370$ & 2018-11-09 09:53:28 & zaf\_0067\_peu & $1.0$ & $3.118039$ & $3.118039$ \\
... & ... & ... & ... & ... & ... & ... & ... \\
217 & $158.760289$ &	$25.0$ &	$41.410504$ & 	$2.183065$ &	$0.207980$ & 2018-11-09 10:00:20 & zaf\_0067\_peu & $2.0$ & $3.655477$ &	$1.827738$ \\
218 & $129.245156$ &	$25.0$ &	$41.410474$ & 	$2.183081$ &	$0.463096$ & 2018-11-09 10:00:22 & zaf\_0067\_peu & $1.0$ & $3.283648$	& $3.283648$ \\
219 & $118.042007$ & $25.0$ &	$41.410454$	& $2.183111$ & $0.470235$ & 2018-11-09 10:00:23 & zaf\_0067\_peu & NaN & NaN & NaN\\
 \hline
 \hline
\end{tabular}
\end{adjustbox}
\end{table}
 
\section*{Technical Validation}

One could question whether linear interpolation can affect any potential pedestrian mobility analysis. We are in fact adding $7,831$ new GPS locations, which corresponds to an increase of $23.6\%$ in the processed dataset. 

\begin{figure}[t]
     \centering
     \begin{subfigure}[b]{0.49\textwidth}
         \centering
         \includegraphics[width=\textwidth]{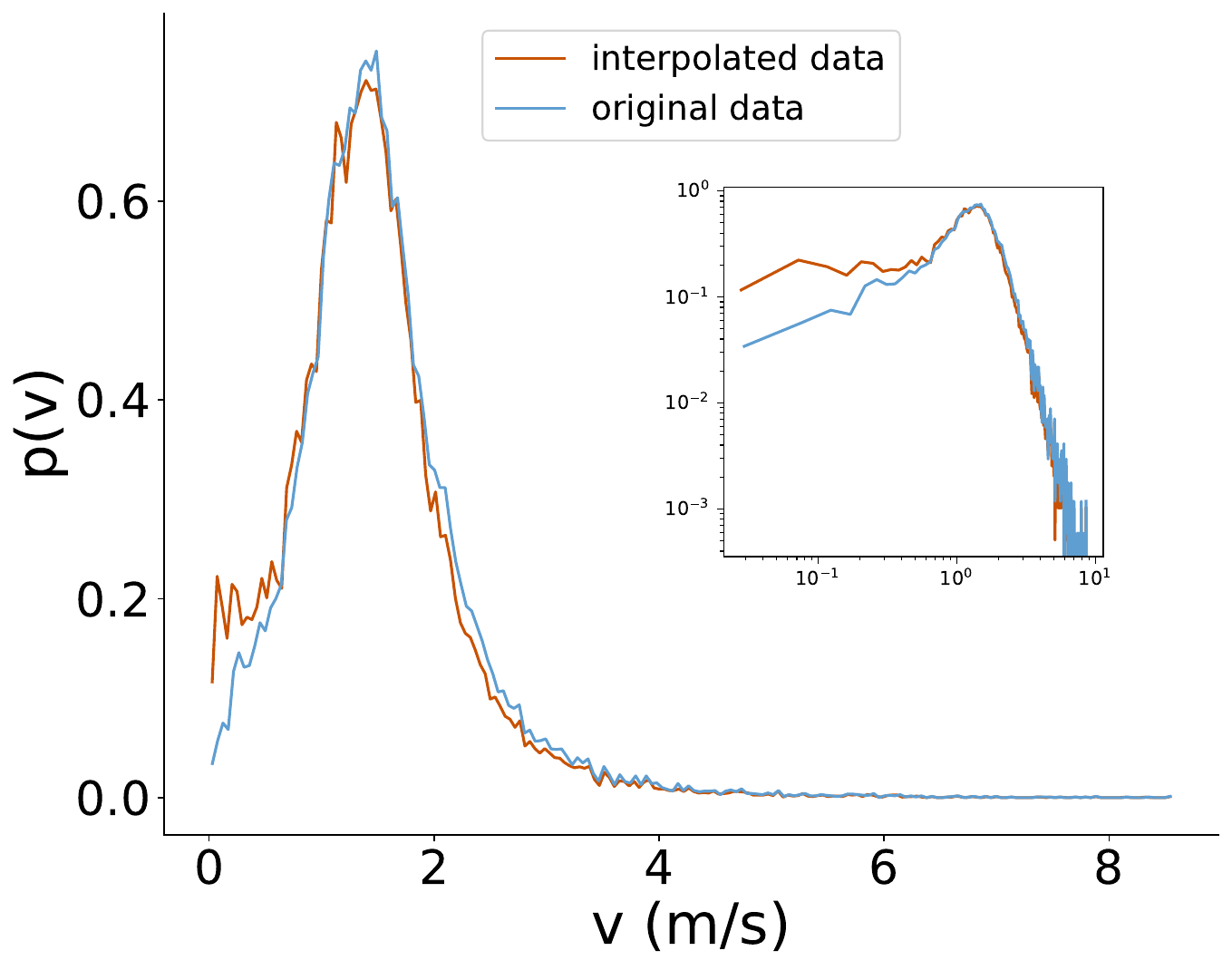}
         \caption{}
         \label{fig:interpolated_figures_v}
     \end{subfigure}
     \begin{subfigure}[b]{0.49\textwidth}
         \centering
         \includegraphics[width=\textwidth]{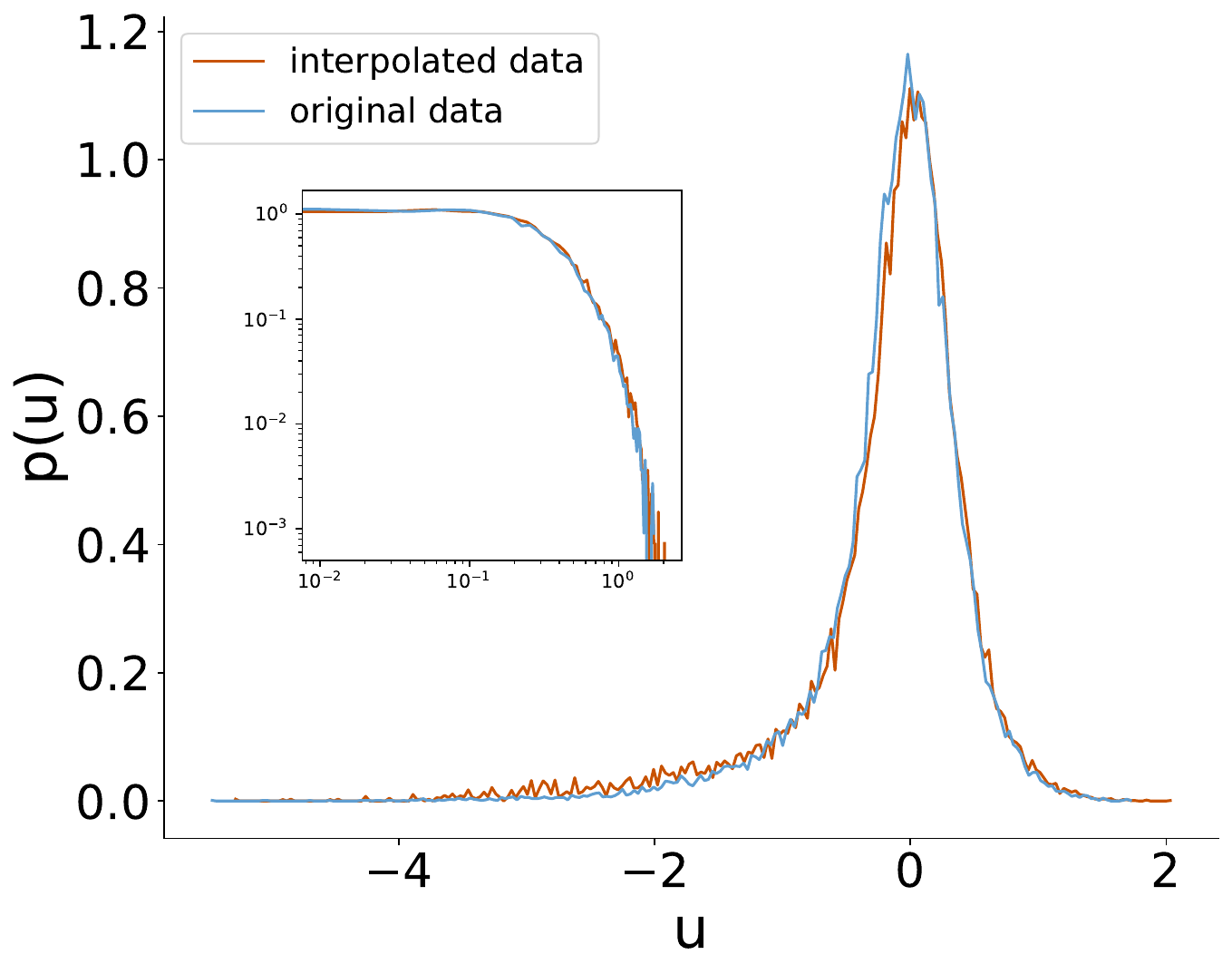}
         \caption{}
         \label{fig:interpolated_figures_u}
     \end{subfigure}
     \begin{subfigure}[b]{0.49\textwidth}
         \centering
         \includegraphics[width=\textwidth]{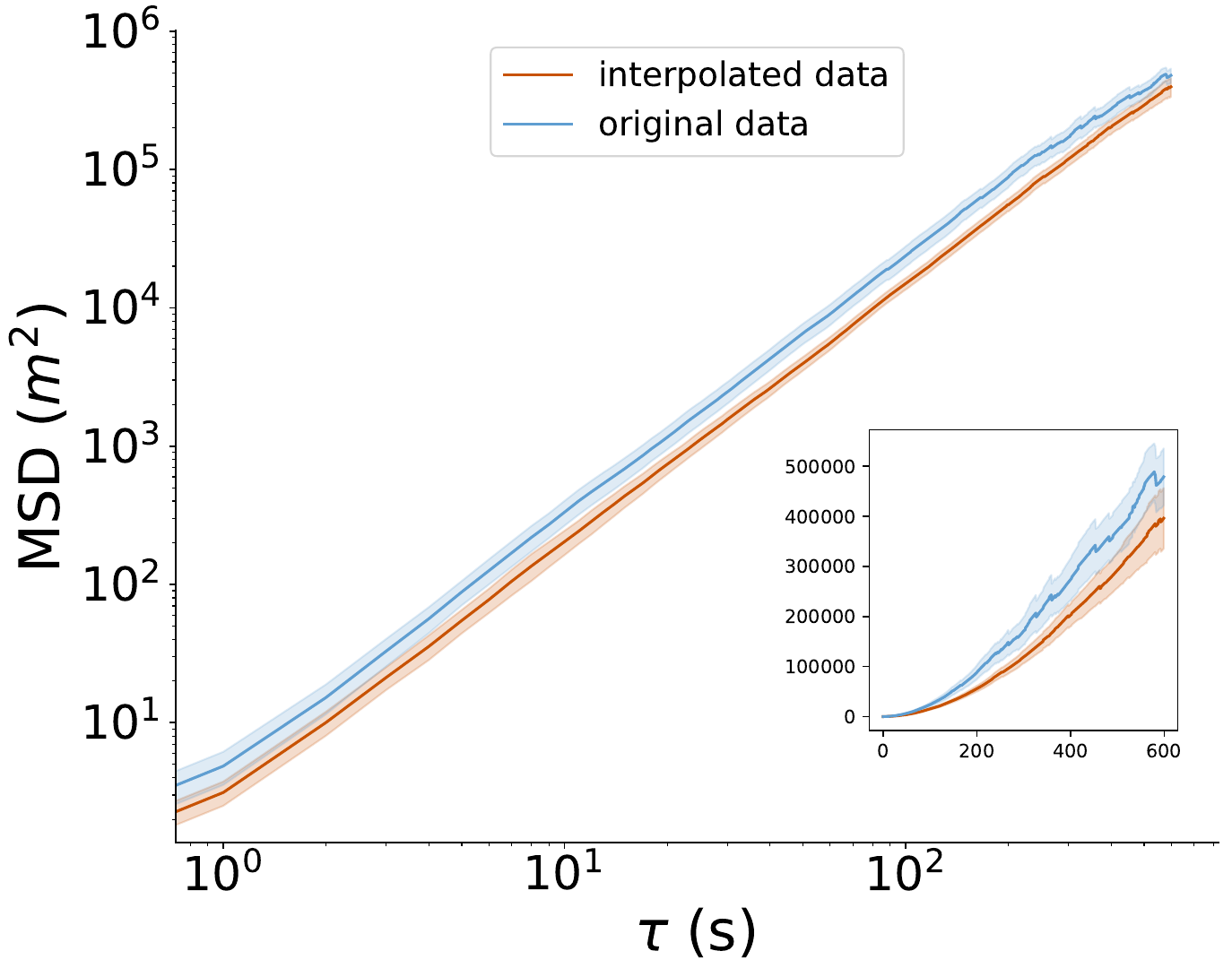}
         \caption{}
         \label{fig:interpolated_figures_msd}
     \end{subfigure}
     \begin{subfigure}[b]{0.49\textwidth}
         \centering
         \includegraphics[width=\textwidth]{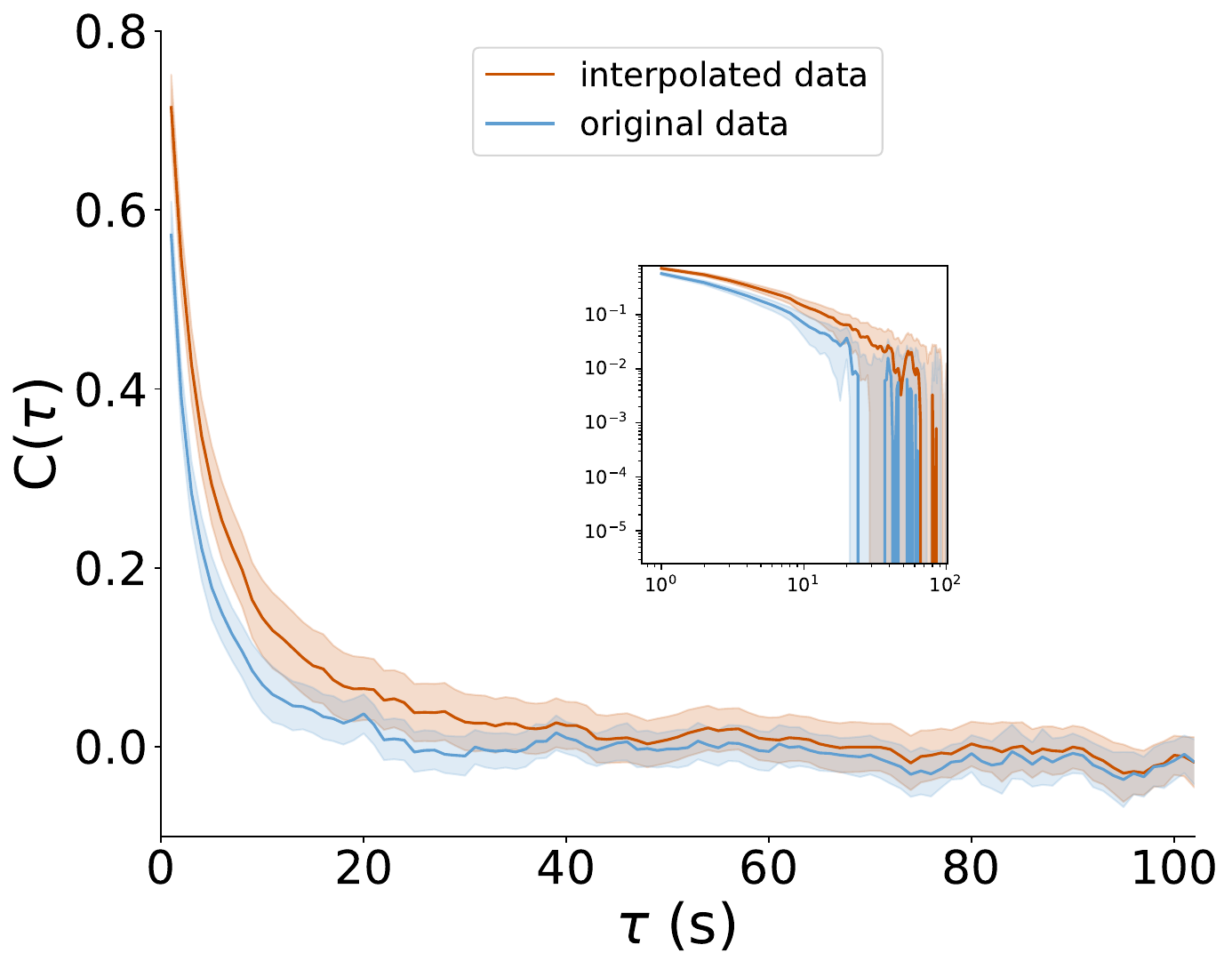}
         \caption{}
         \label{fig:interpolated_figures_autocorr}
     \end{subfigure}
     \caption{ {\bf Main statistical patterns before and after linear interpolation.} (\textbf{a}) Probability density function of instantaneous velocities, $v$ (cf. Eq. (\ref{v})). (\textbf{a}) Probability density function of the logarithm of the instantaneous velocities, $u$ (cf. Eq. (\ref{u})). (\textbf{c}) Mean Squared Displacement with $95\%$ of confidence interval (cf. Eq. (\ref{MSD})). (\textbf{d}) Auto-correlation of the logarithm of the instantaneous velocity with $95\%$ of confidence interval (cf. Eq. (\ref{C})).}
     \label{fig:interpolated_figures}
\end{figure}

Figure \ref{fig:interpolated_figures} compares qualitatively three statistical analyses that can characterize our pedestrian's mobility. First statistical analysis can be made with the probability density $p_V(v)$ of the instantaneous velocity (cf. Eq. (\ref{v})). Another probability density that can also be obtained, $p_U(u)$, can be expressed in terms of the logarithm of the normalized velocity
\begin{equation}
u\equiv \ln(v/v_{m}),
\label{u}
\end{equation}
where $v_m=E[v(t)]$ (for each trajectory). It is also possible to compute the mean squared displacement 
\begin{equation}
MSD(\tau)=E[|\vec{r}(t+\tau)-\vec{r}(t)|^2],
\label{MSD}
\end{equation}
and the auto-correlation of the velocities (cf. Eq. (\ref{u}))
\begin{equation}
C(\tau)=\frac{E[(u(t+\tau)-u_m)(u(t)-u_m)]}{E[(u(t)-u_m)^2]},
\label{C}
\end{equation}
where $u_m=E[u(t)]$. Both quantities are averaged over the trajectories.

Interpolated and non-interpolated data can be thus compared in Figure \ref{fig:interpolated_figures}. Probability densities match qualitatively well except for small velocities. This can be attributed to the interpolation as small time steps are especially relevant in this part of the distribution (see Figure \ref{fig:interpolated_figures_v}). When the logarithm of velocities is considered in Figure  \ref{fig:interpolated_figures_u}, this effect is blurred out. When considering the MSD, the growth with time $\tau$ is shifted and the curves are almost parallel in a double logarithmic scale (see Figure \ref{fig:interpolated_figures_msd}). Figure \ref{fig:interpolated_figures_autocorr} is particularly sensitive to data interpolation as the interpolation enhances correlation as expected, particularly for short time distances. The computed observables, such as the distance traveled (cf. Eq. (\ref{D})) and the time spent (cf. Eq. (\ref{T})) remain statistically similar (see Table \ref{tab:statistics}). Furthermore, the average effective speed $v_{\rm eff}=D/T$ is also analyzed in Table \ref{tab:statistics2}. The mean instantaneous velocity seems to slightly decrease after the linear interpolation showing a shifting effect which was also observed in Figure \ref{fig:interpolated_figures_msd}. The analysis confirms that linear interpolation does not broadly modify the  statistical features here reported. Modifications are subtle but in any case limited and small.

\begin{table}[t]
\centering
\begin{tabular}{ccccc}
\hline \hline
& $T$ (s) & $T$ int. (s) & $D$ (m) & $D$ int. (m) \\ \hline
$\langle \dots \rangle$ & $494\pm 31$ & $494\pm 31$ & $701\pm 40$ & $701\pm 40$ \\
$\sigma$ & $282$ & $282$ & $368$ & $368$ \\
Q1 ($25\%$) & $300$ & $300$ & $467$ & $467$ \\
Q2 ($50\%$) & $462$ & $462$ & $639$ & $639$ \\
Q3 ($75\%$) & $596$ & $596$ & $917$ & $917$\\
min. & $67$ & $67$ & $84$ & $84$ \\
max. & $1,582$ & $1,582$ & $2,305$ & $2,305$ \\
\hline\hline
\end{tabular}
\caption{\label{tab:statistics} {\bf Duration of the trajectories and distance travelled before and after linear interpolation procedure (cf. Eqs. (\ref{T}) and (\ref{D})).} From all participants: mean value ($\langle \dots \rangle$), standard deviation, quantiles, and minimum and maximum value of the duration of the trajectory and the distance travelled before and after linear interpolation.}
\end{table}

\begin{table}[t]
\centering
\begin{adjustbox}{width=\textwidth}
\begin{tabular}{lcccc}
\hline \hline
& $v_{\rm eff}$ (m/s) & $v_{\rm eff}$ int. (m/s) & $v$ (m/s) & $v$ int. (m/s) \\ \hline
$\langle \dots \rangle$ & $1.46\pm 0.03$ & $1.46\pm 0.03$ & $1.519\pm 0.004$ & $1.420\pm 0.004$ \\
$\sigma$ & $0.251$ & $0.251$ & $0.749$ & $0.751$ \\
Q1 ($25\%$) & $1.338$ & $1.338$ & $1.071$ & $0.983$ \\
Q2 ($50\%$) & $1.469$ & $1.469$ & $1.435$ & $1.374$ \\
Q3 ($75\%$) & $1.640$ & $1.640$ & $1.843$ & $1.762$\\
min & $0.864$ & $0.864$& $0.006$ & $0.006$ \\
max & $1.952$ & $1.952$ & $8.566$ & $8.566$ \\
\hline\hline
\end{tabular}
\end{adjustbox}
\caption{\label{tab:statistics2} {\bf Effective speed and instantaneous velocity before and after linear interpolation procedure (cf. Eqs. (\ref{v})).} From all participants: mean value ($\langle \dots \rangle$), standard deviation, quantiles, and minimum and maximum value of the effective speed (defined as the total distance travelled over the time spent, $D/T$) and the instantaneous velocity before and after linear interpolation.}
\end{table}

\section*{Usage Notes}

The collected data from the Beepath citizen science experiment represents a rich source of mobility information for the study of people's micro-behavior around neighborhoods. The data reports trajectories with a specific origin and destination (home-to-school, or the other way around) from an age-uniform group of participants.

Clean pedestrian mobility data can be accessed through the two folders in the repository ({\tt processed data} to processed and clean data and {\tt interpolated data} adding a further step where the data is interpolated and all GPS records are uniformly spaced every 1 second). 

\begin{figure}[t] 
     \centering
     \begin{subfigure}[b]{0.275\textwidth}
         \centering
         \includegraphics[width=\textwidth]{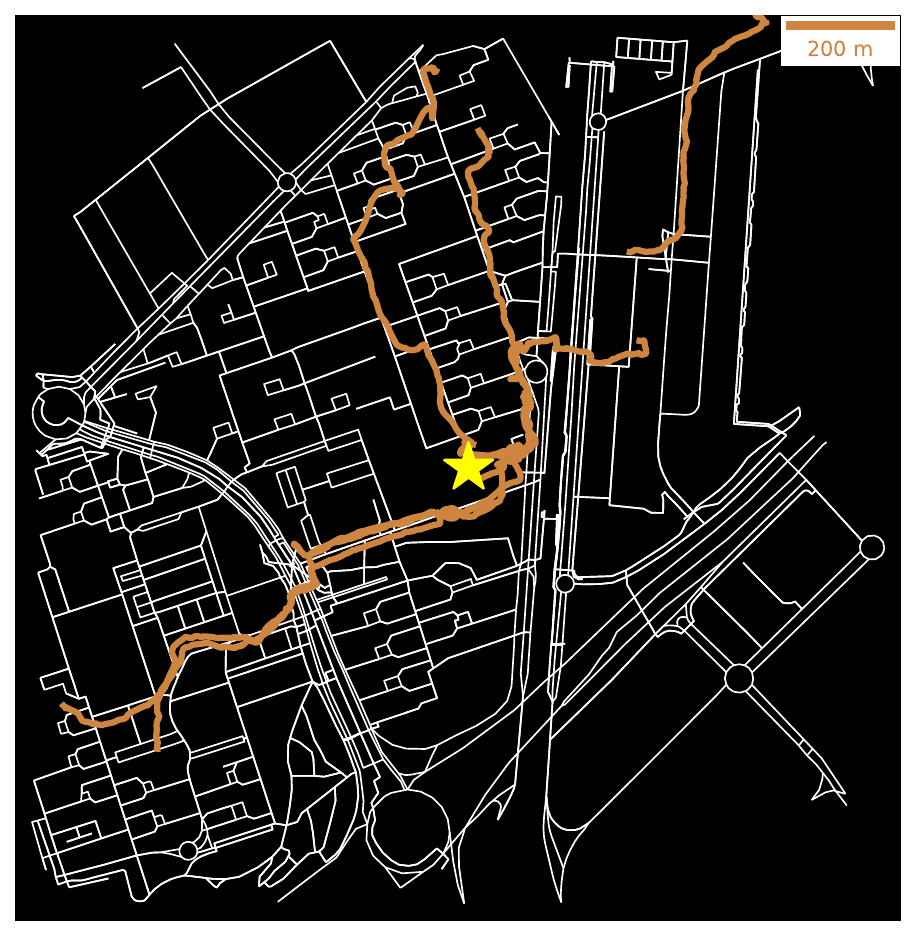}
         \caption{}
         \label{fig:bel_processed}
     \end{subfigure}
     \begin{subfigure}[b]{0.28\textwidth}
         \centering
         \includegraphics[width=\textwidth]{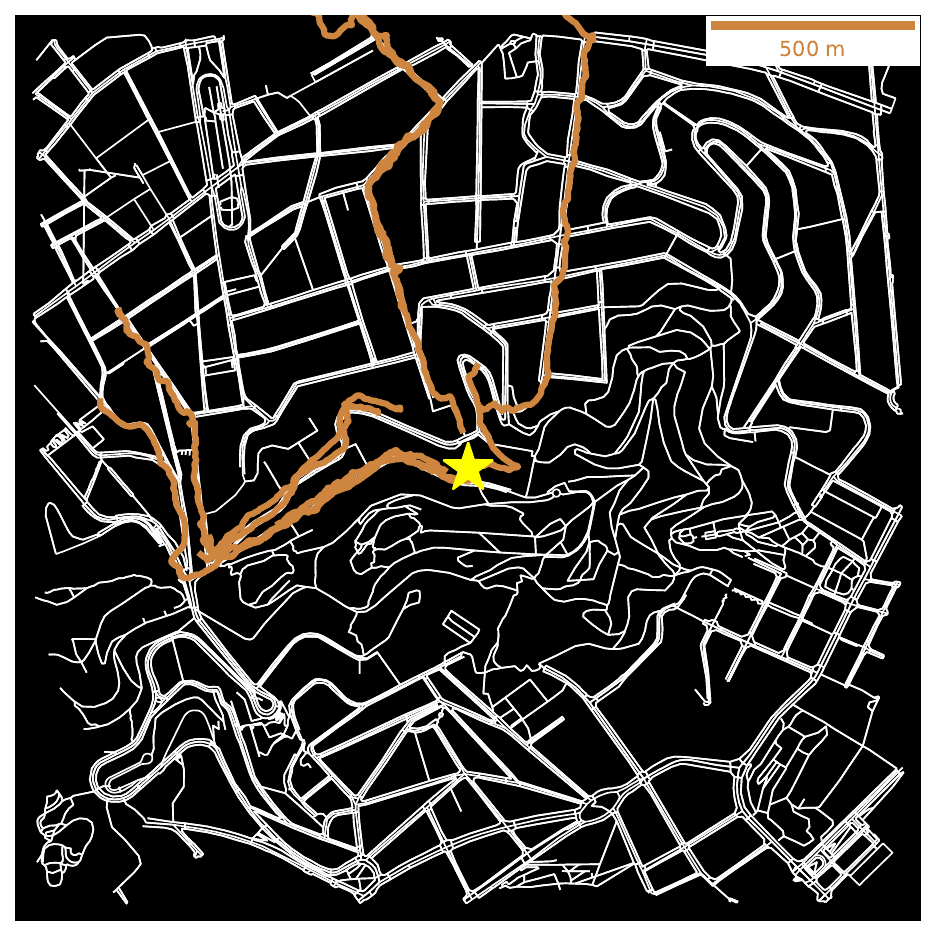}
         \caption{}
         \label{fig:ift_processed}
     \end{subfigure}
     \begin{subfigure}[b]{0.28\textwidth}
         \centering
         \includegraphics[width=\textwidth]{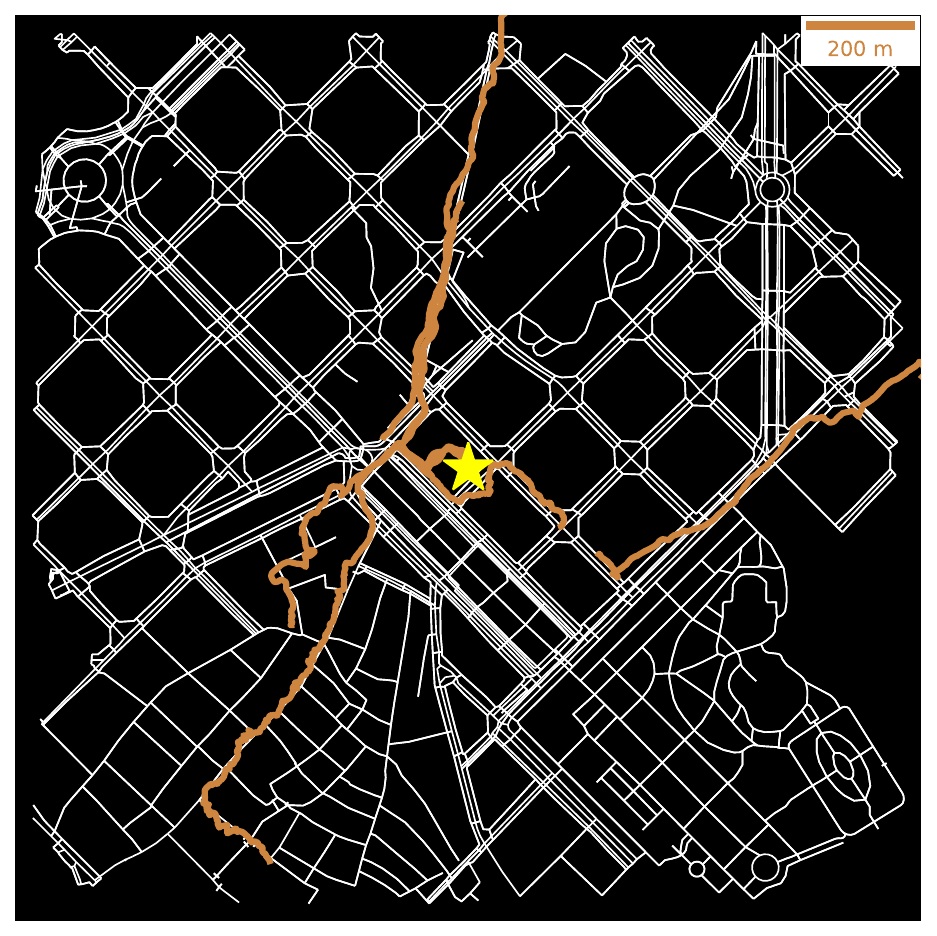}
         \caption{}
         \label{fig:ipc_processed}
     \end{subfigure}
     \begin{subfigure}[b]{0.28\textwidth}
         \centering
         \includegraphics[width=\textwidth]{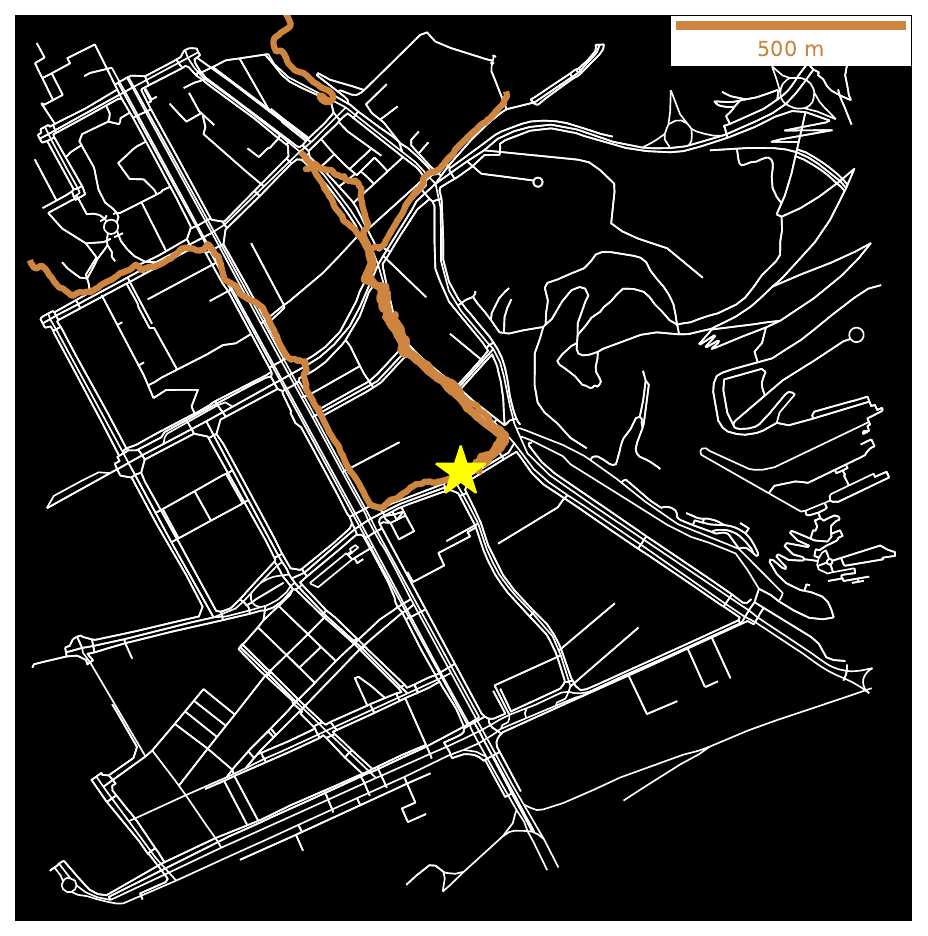}
         \caption{}
         \label{fig:mon_processed}
     \end{subfigure}
     \begin{subfigure}[b]{0.28\textwidth}
         \centering
         \includegraphics[width=\textwidth]{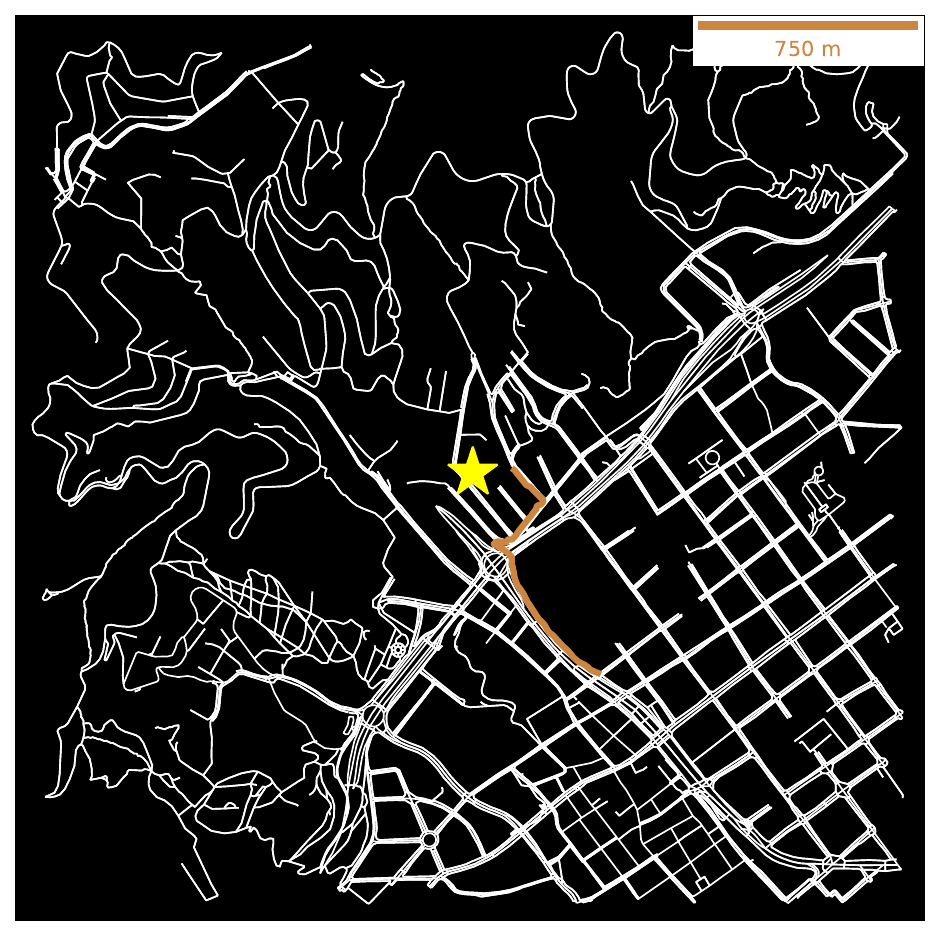}
         \caption{}
         \label{fig:oak_processed}
     \end{subfigure}
     \begin{subfigure}[b]{0.28\textwidth}
         \centering
         \includegraphics[width=\textwidth]{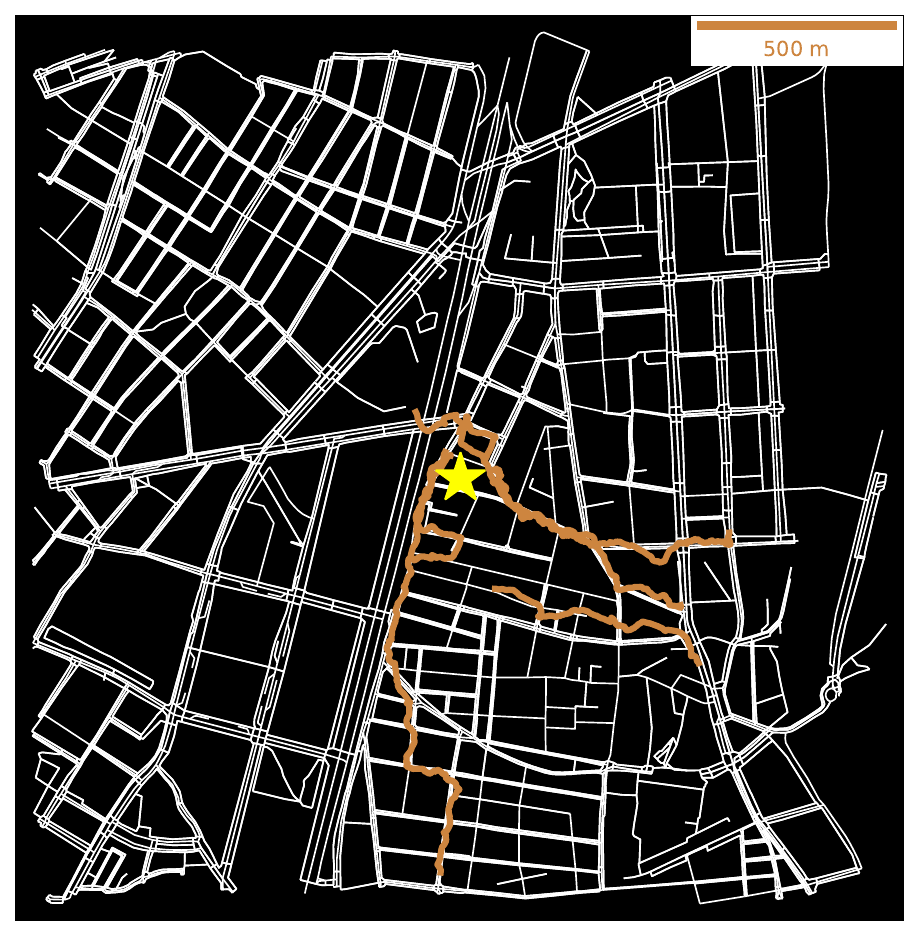}
         \caption{}
         \label{fig:san_processed}
     \end{subfigure}
     \begin{subfigure}[b]{0.30\textwidth}
         \centering
         \includegraphics[width=\textwidth]{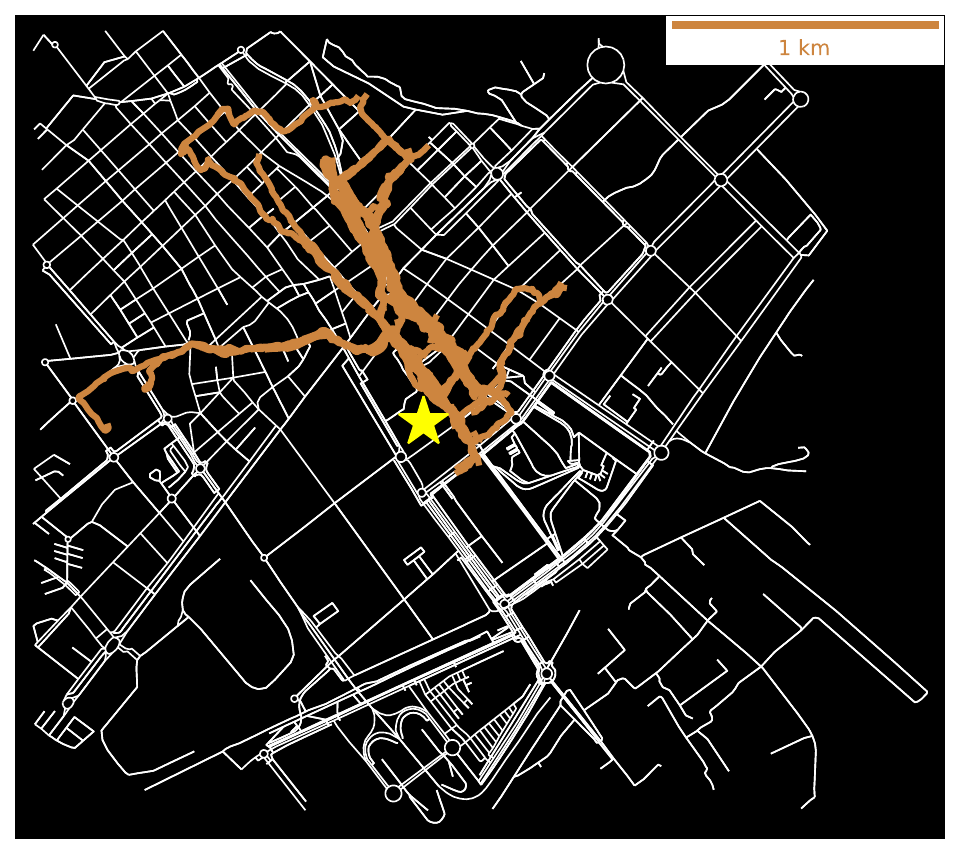}
         \caption{}
         \label{fig:sgv_processed}
     \end{subfigure}
     \begin{subfigure}[b]{0.27\textwidth}
         \centering
         \includegraphics[width=\textwidth]{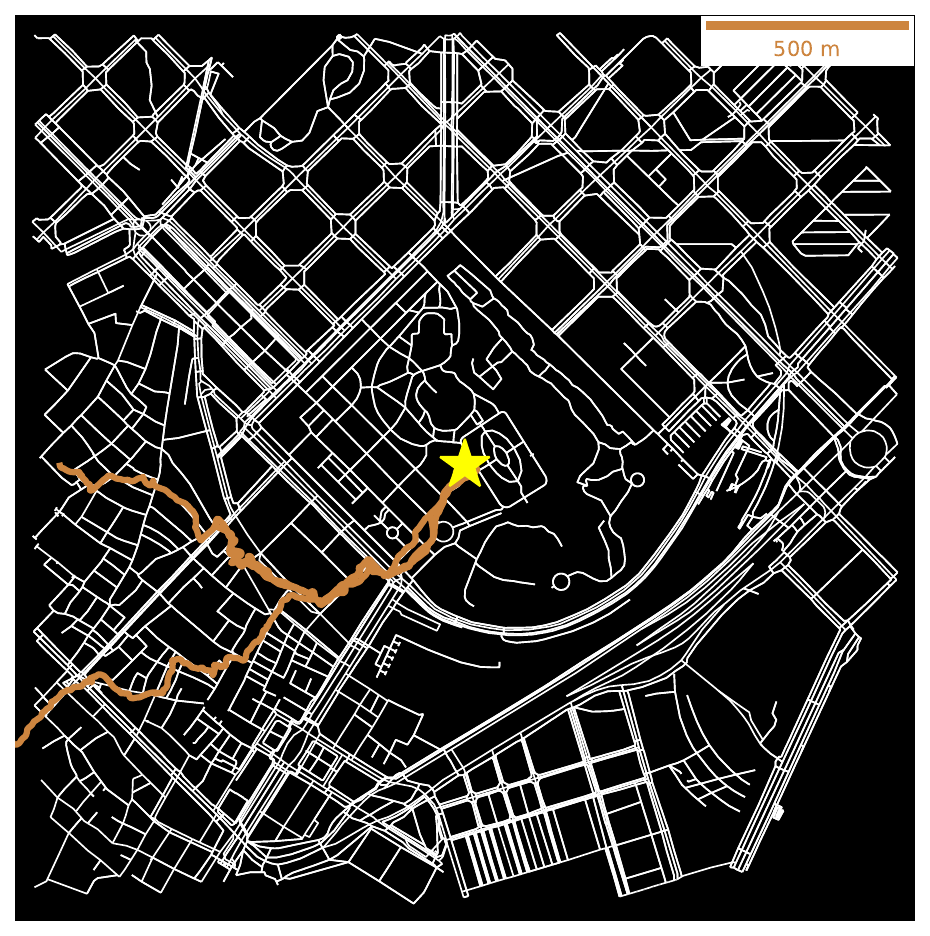}
         \caption{}
         \label{fig:ver_processed}
     \end{subfigure}
     \begin{subfigure}[b]{0.28\textwidth}
         \centering
         \includegraphics[width=\textwidth]{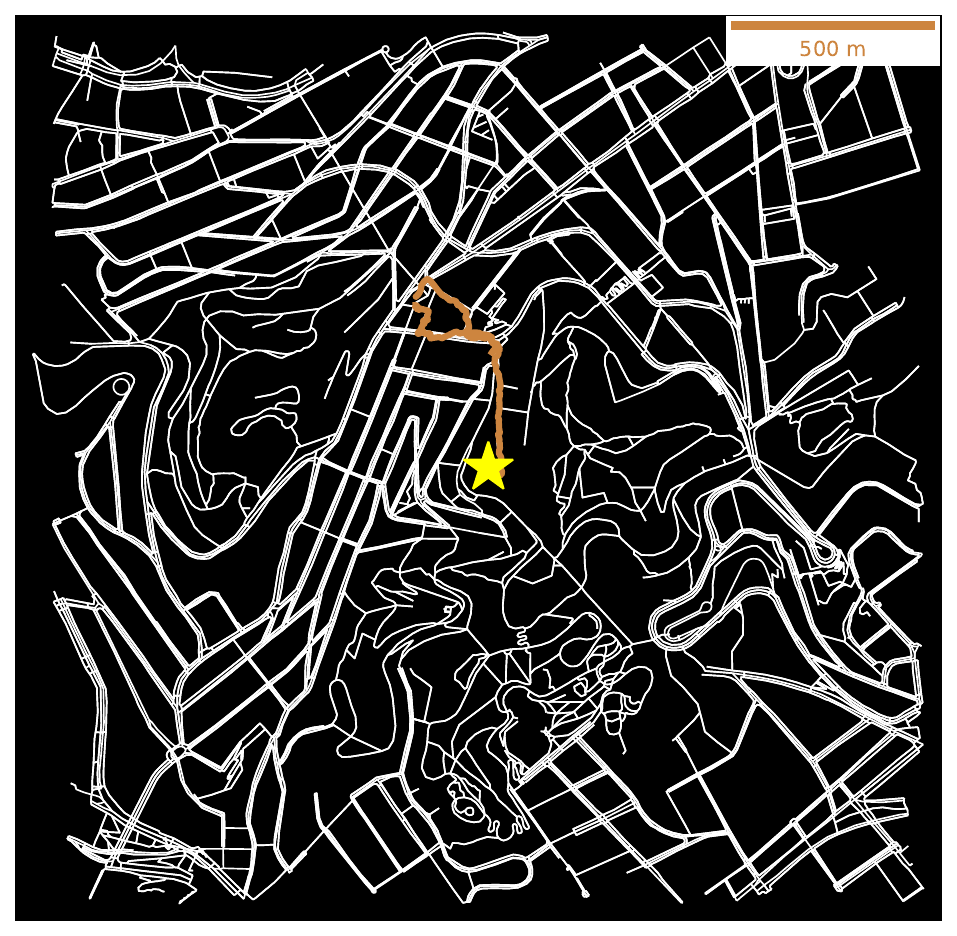}
         \caption{}
         \label{fig:vir_processed}
     \end{subfigure}
     \begin{subfigure}[b]{0.29\textwidth}
         \centering
         \includegraphics[width=\textwidth]{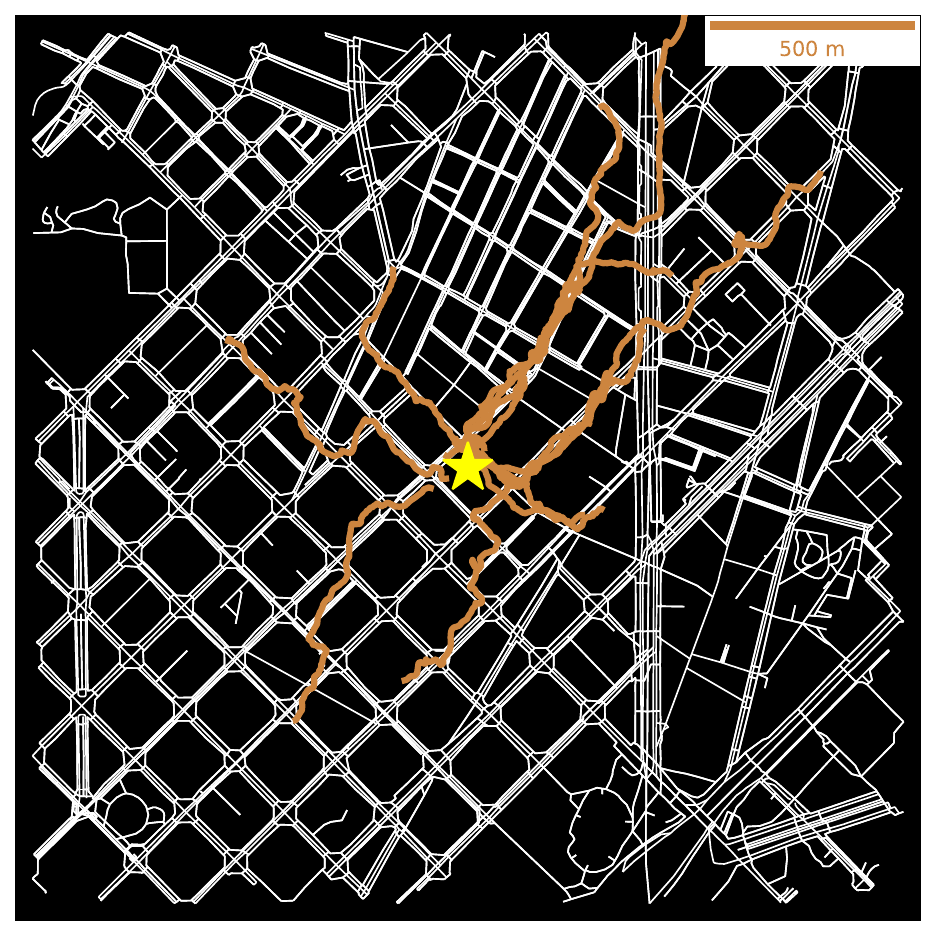}
         \caption{}
         \label{fig:zaf_processed}
     \end{subfigure}
     \caption{{\bf Processed and cleaned trajectories projected on maps for each school.} (\textbf{a}) BEL. (\textbf{b}) IFT. (\textbf{c}) IPC. (\textbf{d}) MON. (\textbf{e}) OAK. (\textbf{f}) SAN. (\textbf{g}) SGV. (\textbf{h}) VER. (\textbf{i}) VIR. (\textbf{j}) ZAF. Yellow star corresponds to school locations.}
     \label{fig:processed_trajectories}
\end{figure}

The Python notebook called {\tt Data Processing.ipynb} in the GitHub repository (\url{https://github.com/ferranlarroyaub/Beepath-Schools.git}) contains a description of the treatment and data clean-up discussed here, as well as the scripts to reproduce all the trajectory representation on maps (Figure \ref{fig:processed_trajectories}). Each processed individual trajectory is saved in a new $csv$ file also contained in the {\tt processed data} folder and in the {\tt interpolated data} folder after performing the linear interpolation. In addition, the Python Notebook also contains the code for the calculation of the time increment $\Delta(t)$ between consecutive timestamps, distances ($d(t)$ and $D$, cf. Eq. (\ref{d}) and Eq. (\ref{D})) and velocities ($v(t)$, cf. Eq. (\ref{v})) between consecutive timestamps. These variables are added as new columns to the processed $csv$ file of each participant. 

The repository (\url{https://github.com/ferranlarroyaub/Beepath-Schools.git}) also contains different Jupyter notebooks with the necessary functions and scripts to study and characterize the participants' movement through statistical patterns. All these files contain a detailed description of the study and an explanation of the code.

In particular, the notebooks called {\tt Mean Squared Displacement.ipynb}, {\tt Instantaneous velocity.ipynb}, and {\tt Autocorrelation velocities.ipynb} contains the scripts to reproduce Figure \ref{fig:interpolated_figures}, where the three main statistical features studied are compared with and without linear interpolation for technical validation purposes. These statistical patterns are the mean squared displacement, the probability density function of the instantaneous velocity, and the auto-correlation of the velocities.

\section*{Acknowledgements}

We acknowledge the participation of more than 427 volunteers and 31 teachers from the schools: OAK House School, Institut Verdaguer, Escola Virolai, Col$\cdot$legi Sagrada Fam\'ilia Sant Andreu, Institut Pau Claris, Institut Bellvitge, Institut Montju\"{i}c, Institut Juan Manuel Zafra, Institut Ferran Tallada, and Col$\cdot$legi Sant Gabriel de Viladecans. We also thank the Consorci d'Educació de Barcelona and the Barcelona City Council through its Citizen Science Office for their commitment to the project and its support to citizen science practices in the city. This work was partially supported by MINEICO (Spain), Agencia Estatal de Investigación (AEI) and Fondo Europeo de Desarrollo Regional (FEDER) [grant number FIS2016-78904-C3-2-P, JP; grant number PID2019-106811GB-C33 (AEI/10.13039/501100011033), JP and FL; grant number PID2019-106811GB-C32 (AEI/10.13039/501100011033), EM]; by Generalitat de Catalunya (Spain) through Complexity Lab Barcelona [grant number: 2017 SGR 608; FL and JP]; and by BarcelonActiva (Impulsem el que fas 2017) [OD, OS, and PC].

\section*{Author contributions}

Ferran Larroya - data validation - writing - proofreading. Ofelia D\'iaz - data acquisition - project conception. Pol Colomer Sim\'on - data acquisition - data validation - project conception. Oleguer Sagarra - data acquisition – data validation - project conception. Salva Ferr\'e - project conception. Esteban Moro - proofreading - writing. Josep Perell\'o - data acquisition - project conception - proofreading - writing.

\section*{Competing interests}

The authors declare that they have no known competing financial interests or personal relationships that could have appeared to influence the work reported in this paper.

%\section*{Figures and figures legends}

%Figure should be referred to using a consistent numbering scheme through the entire Data Descriptor. For initial submissions, authors may choose to supply this document as a single PDF with embedded figures, but separate figure image files must be provided for revisions and accepted manuscripts. In most cases, a Data Descriptor should not contain more than three figures, but more may be allowed when needed. We discourage the inclusion of figures in the Supplementary Information \textendash{} all key figures should be included here in the main Figure section. 

%Figure legends begin with a brief title sentence for the whole figure and continue with a short description of what is shown in each panel, as well as explaining any symbols used. Legend must total no more than 350 words, and may contain literature references. 

%\section*{Tables}

%Authors are encouraged to provide one or more tables that provide basic information on the main ‘inputs’ to the study (e.g. samples, participants, or information sources) and the main data outputs of the study; also see the additional information on providing metadata on page 6. Tables in the manuscript should generally not be used to present primary data (i.e. measurements). Tables containing primary data should be submitted to an appropriate data repository.

%\subsection*{Citing Data}
%In line with emerging industry-wide standards for data citation, references to all datasets described or used in the manuscript should be cited in the text with a superscript number and listed in the ‘References’ section in the same manner as a conventional literature reference. See the examples above.

\end{document}